\documentclass[11pt]{article}

\usepackage[utf8]{inputenc}
\usepackage[T1]{fontenc}
\usepackage{lmodern}

\usepackage{amsmath,amssymb}
\usepackage{mathtools}   
\usepackage{bm}          
\usepackage{amsthm}      

\usepackage{booktabs}    
\usepackage{array}       
\usepackage{multirow}    
\usepackage{tabularx}    
\usepackage{longtable}   

\usepackage{graphicx}
\usepackage{epstopdf}    
\usepackage{subcaption}
\usepackage[labelfont=bf,textfont=it]{caption}
\graphicspath{{Figures/}} 

\usepackage{geometry}
\usepackage{setspace}
\usepackage{multicol}
\usepackage{enumitem}
\usepackage[section]{placeins} 

\usepackage{microtype}  

\usepackage{cite}
\usepackage{xcolor}
\usepackage[colorlinks=true,
linkcolor=blue,
citecolor=blue,
urlcolor=blue]{hyperref}
\usepackage{cleveref}   

\hypersetup{
	pdftitle={Electroosmotic lubrication flow in constricted microchannels with a compliant wall and DLVO interactions},
	pdfauthor={Subhajyoti Sahoo and Ameeya Kumar Nayak},
	pdfsubject={Electroosmotic lubrication in compliant microchannels},
	pdfkeywords={solute dispersion,
		multiple charged species, lubrication approximation,
		low-Reynolds-number flows,
		separation efficiency.}
}

\usepackage{authblk}

\title{Dispersion of multiple charged species in an axially symmetric slowly varying channel}
\author[1]{Thakurdas Mahata}
\author[2]{Anirban Chatterjee}
\author[1]{Ameeya Kumar Nayak\thanks{Corresponding author: \href{mailto:ameeyakumar@gmail.com}{ameeyakumar@gmail.com}}}
\affil[1]{Department of Mathematics, IIT Roorkee, India, Uttarakhand, Roorkee, Pin-247667}
\affil[2]{Univ. Bordeaux, CNRS, LOMA, UMR 5798, F-33405 Talence, France}
\date{}
\begin{document}
		
	\maketitle
	
	\begin{abstract}
			The transport and dispersion of multiple species of charged ions are central to many biological and physical processes, including electrokinetic ion separation. However, most theoretical studies of dispersion in channels have focused on neutral solutes, leaving the transport of multiple charged species comparatively unexplored. Differences in ionic diffusivities in a multispecies electrolyte solution generate an self-induced electric fields that drive electromigration. To capture these effects at the macroscopic scale, we combine the lubrication approximation with homogenization theory, under electroneutrality and zero-current constraints, to derive an effective transport equation governing the cross-sectionally averaged concentrations. We apply our model framework to a range of channel geometries and compute the resulting effective dispersion coefficients. Finally, we investigate how channel geometry can be tuned to enhance ionic separation. We observe a geometry-induced electro-diffusive coupling that inhibits solute dispersion in certain channels, leading to a
			non-monotonic Number of Theoretical Plates (NTP) and making such channels ideal for separation processes.

			\vspace{0.5cm} 
			\end{abstract}

	\maketitle
	\section{Introduction}	
	Transport of ions associated with multicomponent diffusion has attracted significant attention in microfluidic systems and is considered a potent mechanism for separating species in biological and chemical processes. The concentrations of ionic species, their charges, and their mobilities play key roles in selecting the optimal transport methods. The study of particle transport in fluid flows plays a crucial role in biomedical devices \cite{fogelson2015fluid}, contributing to the advancement in mixing of biological species \cite{wang2019finding}, diffusing \cite{kamholz2001theoretical}, and separating \cite{pamme2007continuous} colloidal systems and biological cells, including the cancer diagnostics and blood analysis \cite{guo2021multifunctional, li2018paper}. The interaction of fluid flow and molecular diffusion causes solutes to spread, a phenomenon known as the Taylor-Aris dispersion, which is widely used to determine the diffusion coefficients of chemical species \cite{d2008determination} and to enhance mixing in microchannel flows \cite{stone2004engineering}. The theory was first developed by Taylor \cite{taylor1953dispersion} to study the dispersion of passive solutes and developed a closed-form solution for the long-term dispersive behaviour of solutes in a circular cylindrical tube where the flow was assumed to be a fully developed Poiseuille flow. Later on, the study was further extended for a wider range of Peclet numbers using the statistical method of moments \cite{aris1956dispersion}. Stone \& Brenner \cite{stone1999dispersion} study the Taylor's dispersion for slow varying fluid velocity along the axial direction. Scaling laws and numerical approaches were used to explore various mechanisms associated with each dispersion regime \cite{lighthill1966initial, chatwin1970approach}. Classical Taylor dispersion theory enables the reduction of the dimension by approximating the governing equations with a smaller number of independent variables \cite{young1991shear, teng2023diffusioosmotic, guan2024streamwise, chu2019dispersion}. However, the solutes that require microfluidic manipulation carry an intrinsic charge, and the methods used for studying Taylor-Aris dispersion fail to capture the mechanisms associated with charged solutes. \\	
	Taylor dispersion of charged species describes the dispersion of solutes, in which molecular diffusion couples with the flow and the electric field. In an electrolyte solution, the electric current is carried by the dissociated ions. The electric field exerts significant body forces on the ions, affecting their fluxes. The dispersion of ionic species in microchannels is largely charge dependent and significantly deviates from that of neutral solutes in the same flow \cite{de2006ionic}. The presence of electromigration interactions leads to distinct behaviour in charged species, influencing their dispersion and movement in a medium, unlike neutral molecules \cite{ghosal2012electromigration, mikkers1979high, chatterjee2022effect, chatterjee2024effect}. Even in the absence of an external field, the differences in diffusivity of multiple charged species lead to the generation of an induced electric potential \cite{ding2023shear}, which transports the ionic component. In binary electrolyte systems, the flux of any ionic species is directly proportional to the gradient of concentration of that particular ionic species. However, in a multi-component ionic species, the flux of any species is non-linearly coupled with the concentration gradient of the other ionic species \cite{gupta2019diffusion}, leading to a non-Fickian diffusive dynamics.\\
	In general, it is observed that the flow geometry is crucial in determining the extent of Taylor dispersion, because variations in cross-sectional shape directly influence the velocity profile and shear rate of the flow \cite{xu2013upscaling}. The optimised complex geometries, such as microconduits with periodically varying cross-sections, used for strategic mixing phenomena require the movement and manipulation of species \cite{haugerud2022solute, kalinay2020taylor, roggeveen2023transport}. Hogland \& Prud'Homme \cite{hoagland1985taylor} applied the framework of Frankel \& Brenner \cite{frankel1989foundations} to study the solute dispersion in sinusoidal channels using the method of moments. Rosencrans \cite{rosencrans1997taylor} studied Taylor dispersion in curved channels and concluded the existence of some channel shapes and flow fields that lead to a reduction instead of an increase in solute dispersion. The method of local moments was used to study the impact of the geometry and molecular diffusion of effective solute dispersion analytically \cite{bolster2009solute}, and it was found that for a fixed flow rate, periodic fluctuations of the channel aperture can lead to both an increase and a decrease in effective dispersion depending on the fluctuation ratio. Recently, Chan \& Santiago \cite{chang2023taylor} analysed the Taylor–Aris dispersion in an arbitrarily shaped axisymmetric channel using the method of moments and derived simplified equations for solute evolution. Bryden \& Brenner \cite{bryden1996multiple} studied Taylor-Aris dispersion within geometries such as a diverging conical channel, flared and axisymmetric Venturi tube. Multiple time scale analysis was also used to derive an asymptotic expression for the effective dispersion coefficient. The approach  was centred around developing a reduced order partial differential equation  to describe the conditional probability density function associated with the solute transport.\\
	However, in the transport of biological fluids 
	\cite{wang2025microfluidic}, 
	fluid flow in energy storage devices \cite{gupta2020charging}, and in many other phenomena, ionic charges are of significance, and the role of ionic mobility cannot be ignored.
	In many of these systems, the fluid phases may contain multiple electrolytes with varying physicochemical properties. Several studies have discussed the transport of multiple charged species in industrial applications, such as in cement pores \cite{johannesson2007multi} and in electrochemical processes \cite{tsaoulidis2015effect, sprocati2020charge}. In studies concerning multiple charged ionic species composed of two salts with a common cation, it has been observed that ion fluxes are coupled and depend on the relative diffusivities of the ions, the ratio of the ionic concentration in pores and reservoirs, and the valencies of the different species \cite{gupta2019diffusion}. Ding \cite{ding2023shear} developed a framework to analyze the Taylor-Aris dispersion for the multispecies electrolyte solution in a straight channel where the fluid is driven by a Poiseuille flow and noted that the differences in ion diffusivities induced an electric potential and  generate additional fluxes for each species. \\
	To the best of our knowledge, two disparate branches of research have been widely pursued, one focusing on the role of geometric modulation in Taylor-Aris dispersion, while the other focuses on the dispersion of multiple species in straight channels. Analysing the effect of wall slope on the evolution of charged species has significant potential to influence the design and analysis of a wide range of systems, including microfluidic devices, yet the role of the convexity-concavity of channel geometry in the transport of charged ionic species remains unexplored. The present framework aims to address this research gap, and we study the Poiseuille flow-driven dispersion of multiple charged ionic species within an arbitrary, yet slowly varying, microchannel. We employ homogenization methods to derive a macrotransport equation for the evolution of the transversely averaged species concentrations at long times.\\
	The organization of the paper is as follows. Section $\ref{sec2}$ presents the model description and the assumptions utilized in describing our model. Section $\ref{sec3}$ presents the detailed derivation of the reduced-order macrotransport equation in the long-time limit using a multiple time-scale approach. Numerical validation of our results by comparison with those reported by \cite{ding2023shear}, is provided in section \ref{sec4}. In Section \ref{sec5}, we consider several channel geometries, and present the variations of the transverse-averaged concentrations and the corresponding effective diffusivities, and identify an optimal geometry that maximizes separation efficiency for the charged species. Finally, in section 6, we conclude with a summary of results and discuss potential avenues for future research.
	
	\section{Model Description} \label{sec2}
	
	\begin{figure}
		\centering
		\includegraphics[width=0.8\textwidth]{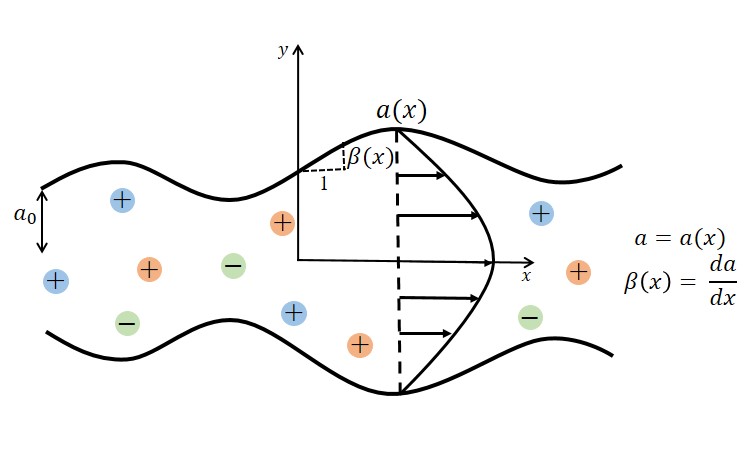}
		\caption{Schematic diagram of dispersion of multiple charged species under an induced electric field, driven by a Poiseuille flow.}
		\label{fig1}
	\end{figure}
	In this model the transport of multispecies electrolyte solution in the presence of Poiseuille flow is considered in an axisymmetric channel domain: $(x, y)\in \mathbb{R} \times \Omega$, with a slowly varying arbitrary radius $a=a(x)$, where $\Omega = \{y:y\in[-a(x), a(x)]\}$, and the $x$-direction is the longitudinal direction of the channel. Here, ${\textbf{n}}$ is the outward normal vector of the boundary $\mathbb{R} \times \partial \Omega$, where $\partial \Omega$ is the boundary of $\Omega$.
	The variables for the first and second derivatives are denoted as $\beta(x)=\frac{da}{dx}$ ,and  where $\gamma(x)= \frac{d^2a}{dx^2}$.
	\subsection{Equations of fluid flow}
	The fluid velocity, $\mathbf{u}$, and pressure p, in the region $\Omega$ representing the interior of the channel are governed by the incompressible Navier-Stokes flow equation and the continuity equation:
	\begin{align}
		\rho( \frac{\partial \mathbf{u}}{\partial t} + \mathbf{u}\cdot\nabla\mathbf{u} ) = -\nabla p + \mu\nabla^{2}\mathbf{u} \label{1}
	\end{align}
	\begin{align}
		\nabla\cdot\mathbf{u} = 0, \label{2}
	\end{align}
	where the constants $\rho$ and $\mu$ are, respectively, the density and viscosity of the fluid.
	\subsection{Nernst-Planck equation}
	The electrolyte solution is consider for a system of $N$ ionic species and denoted by ${c}_{i}(x, {y},t)$ the number of ions of the $i^{th}$ species $(i=1,2,...., N)$ present per unit volume and $z_{i}$ is the valency of the $i^{th}$ species, and $t$ is the time. The concentration evolution of $N$ ion species under the shear flow advection and ionic interaction can be modelled by the Nernst-Planck equation and is given by \cite{masliyah}
	\begin{align}
		\frac{\partial{c_{i}}}{\partial{t}} + {\nabla}\cdot(\mathbf{u}c_{i}) = k_{i}\Delta c_{i} + \frac{k_{i}z_{i}e}{k_{B}T}{\nabla}\cdot(c_{i}{\nabla}\phi),\hspace{0.3cm} c_{i}(x, y, 0) = c_{I,i}\left(\frac{x}{\lambda}\right),\hspace{0.2cm} i=1,2,....N, \label{3}
	\end{align}
	where $k_{i}$ is the diffusivity of the $i^{th}$ species of ion, $\phi(x,{y},t)$ is the electric potential, $e$ is the elementary charge, $k_{B}$ is the Boltzmann constant, $T$ is the temperature, $c_{I, i}$ is the initial condition of the $i^{th}$ species of ion, and $\lambda$ is the characteristic wavelength of spatial variation of the channel. The second term on the left-hand side of the above equation describes the fluid flow advection. The first term on the right-hand side of the same equation describes the ion diffusive motion, while the second term represents the electromigration in response to the local electric field.
	\\
	It is assumed that the electrolyte solutions are advected passively by a prescribed velocity field that takes the form $\mathbf{u}=\left(u(x,y), v(x,y)\right)$. The functions $u(x,y)$ and $v(x,y)$ vanish along the boundaries. 
	No-flux boundary condition is imposed for the concentration fields of the ion species
	\begin{align}
		\mathbf{n} \cdot \left(k_{i}{\nabla}c_{i} + (\frac{k_{i}z_{i}e}{k_{B}T})c_{i}{\nabla}\phi\right)\Big|_{\mathbb{R} \times \partial \Omega} = 0, \label{4}
	\end{align}
	where $\mathbf{n} = \frac{\mp\beta(x)\hat{i} + \hat{j}}{\sqrt{1+\beta(x)^2}}$
	is the outward normal to the boundary $y=a(x)$ and $y=-a(x)$ at a fixed value of x respectively.
	\\
	We have system of $N$ species conservation equations for concentration where the electric potential $\phi$ is involved. The electrical potential $\phi$ is governed by the Poisson equation concerning the electric potential which can be easily obtained from Gauss's law \cite{masliyah}. However, this work is focus on the case, where the external electric field is absent. It is assumed that the net charge density is zero almost everywhere. To reduce the complexity of the model it assumed that the solution is electrically neutral, which states that the charge density is zero almost everywhere, and it is mathematically expressed as $\sum\limits_{i=1}^{N}z_{i}c_{i}=0$. The system we have considered in this analysis is devoid of any external electric field and consists of ions having different mobilities. Under the application of a pressure gradient, the ions migrate differentially due to their differences in ionic mobilities, leading to charge separation that drives an electric field and ensures the net electric current in the system is zero. This is expressed by \cite{masliyah}
	\begin{align}
		\mathbf{0} = \sum_{i=1}^{N}z_{i}\mathbf{J}_{i}=\sum_{i=1}^{N}z_{i}\left(\mathbf{u}c_{i}-k_{i}\;{\nabla}c_{i} + (\frac{k_{i}z_{i}e}{k_{B}T})\;c_{i}{\nabla}\phi\right), \label{5}
	\end{align} 
	which is commonly used to conserve the net current in a system when the fluid is not influenced by any externally applied electric field \cite {gupta2019diffusion, boudreau2004multicomponent}. 
	\\
	The gradient of the electric potential can be expressed in terms of ion concentration:
	\begin{align}
		\frac{e}{k_{B}T}\nabla\phi=\frac{\sum\limits_{i=1}^{N}z_{i}(\mathbf{u}c_{i}-k_{i}\:\nabla c_{i})}{\sum\limits_{i=1}^{N}z_{i}^2k_{i}c_{i}} = \frac{\sum\limits_{i=1}^{N-1}z_{i}(k_{n}-k_{i})z_{i}\nabla c_{i}}{\sum\limits_{i=1}^{N-1}(z_{i}k_{i}-z_{n}k_{n})z_{i}c_{i}}, \label{6}
	\end{align}
	where the second equated term is obtained using the electroneutrality condition. This equation clearly explains that the concentration of all of the ions influences the local electric field, which in turn generates the electromigration of ions. Also, the electric potential gradient is induced by the difference in ion diffusivities. When all the diffusivities of different species take the same value, the gradient of the diffusion-induced potential becomes zero, i.e., $\nabla \phi = 0$, then equation (\ref{3}) becomes simply an advection-diffusion equation. If the diffusivities are different, then we can substitute the equation (\ref{6}) in the Nernst-Planck equation (\ref{3}), as  
	\begin{align}
		\frac{\partial{c_{i}}}{\partial{t}} + u(x,{y},t)\frac{\partial{c_{i}}}{\partial{x}} + v(x,{y},t)\frac{\partial{c_{i}}}{\partial{y}} = k_{i}\Delta c_{i} + k_{i}z_{i} {\nabla}\cdot\left(\frac{c_{i}\sum\limits_{j=1}^{N-1}(k_{n}-k_{j})z_{j}\nabla c_{j}}{\sum\limits_{j=1}^{N-1}(z_{j}k_{j}-z_{n}k_{n})z_{j}c_{j}}\right),\nonumber\\ 
		i=1,2,...,N-1, \label{7}
	\end{align}
	The major analysis of this study dealt with equation (\ref{7}), where the induced electric potential $\nabla\phi$ is expressed in terms of the concentration gradient, valency, and diffusivities of the different species and also study the interplay between Poiseuille flow advection and ion-electric interactions in electrically neutral multispecies electrolyte solutions for different channel geometries.
	\subsection{Non-Dimensionalization}
	It is assumed that $a_{0}$ is the inlet height of the channel, which is considered to be the characteristic length along the transverse direction.
	Then, $\epsilon = \frac{a_{0}}{\lambda}$ is a very small parameter of the problem. Here, $U_{0}$ is the averaged axial velocity at the inlet, taken as the characteristic axial velocity. To identify the dominant terms, we have performed  the nondimensionalization of the equation. The change of variables for the nondimensionalization is
	\begin{align}
		\lambda x^{*} = x,\hspace{0.5cm} a_{0} y^{*} = {y}, \hspace{0.5cm} \epsilon=\frac{a_{0}}{\lambda}(<<1), \hspace{0.5cm} \frac{\lambda \epsilon^{2}}{\bar{k}}t^{*} = t, \hspace{0.5cm} a_{0}a^{*}(x^{*})=a(x), \label{8}
	\end{align}
	\begin{align}
		\bar{c}c_{i}^{*}=c_{i},\hspace{0.5cm} \frac{k_{B}T}{e} \phi^{*} = \phi, \hspace{0.5cm} U_{0}u^{*}=u,
		\hspace{0.5cm} \epsilon U_{0}v^{*}=v, \hspace{0.5cm} \frac{\mu U_{0} \lambda}{a_{0}^2}p^{*}=p, \hspace{0.5cm} \epsilon \beta^{*}(x^{*})=\beta(x) \label{9}
	\end{align}
	where $\bar{c}$ is the characteristic concentration, and $\bar{k}$ is the characteristic diffusivity. Here, $v=O(\epsilon)$ and $u>>v$ hence  $p=p(x)$.
	Then, for the steady state solution, after dropping the '$*$' in the flow variables, without loss of generality the nondimensional form of the continuity and Navier-Stokes equations are written as, 
	\begin{align}
		\frac{\partial u}{\partial x} + \frac{\partial v}{\partial y} = 0 \label{10}
	\end{align}
	\begin{align}
		Re \epsilon \left(u\frac{\partial u}{\partial x} + v \frac{\partial u}{\partial y}\right) = - \frac{\partial p}{\partial x} + \epsilon \frac{\partial^{2} u}{\partial x^{2}} + \frac{\partial^{2} u}{\partial y^{2}}\label{11}
	\end{align}
	\begin{align}
		\frac{\partial p}{\partial y}=0,\label{12}
	\end{align}
	where $Re = \frac{\rho a_{0} U_{0}}{\mu}$ is the Reynolds number based on the characteristic channel height. For  $Re<<1$ the above governing set of equations reduces to,
	\begin{align}
		\frac{\partial u}{\partial x} + \frac{\partial v}{\partial y} = 0,
		\hspace{1cm}\frac{\partial^{2}u}{\partial y^{2}}=\frac{dp}{dx}\label{13}
	\end{align}
	along with the boundary condition at $y=\pm a(x)$, $u=0$, $v=0$.
	After solving the nondimensional Navier-Stokes equation using the boundary conditions with the assumption of lubrication theory, we obtain the velocity field along the axial direction as
	\begin{align}
		u(x, y) = \frac{3Q}{4a}\left(1-\frac{y^{2}}{a^{2}}\right)\label{14} + \mathcal{O}(\epsilon^2 Re, \epsilon^2)
	\end{align}
	where $Q$ is the volumetric flow rate and it is given by 
	\begin{align}
		Q=\int_{-a(x)}^{a(x)}u(x,y)dy = -\frac{2}{3}a^3\frac{\partial p}{\partial x}.\label{15}
	\end{align}
	Now using continuity equation along with $\frac{\partial Q}{\partial x}=0$, we obtain, 
	\begin{align}
		v(x, y) = \frac{3Q}{4a} \beta \left(\frac{y}{a}-\frac{y^3}{a^3}\right) + \mathcal{O}(\epsilon^3 Re, \epsilon^3). \label{16}
	\end{align} 
	The reduced non-dimensionalized Nernst-Planck equation is
	\begin{align}
		\frac{ \partial{c_{i}} }{ \partial{t} } +  Pe\;\epsilon \left(  u \frac{ \partial{ c_{i} } }{ \partial{x} } + v \frac{ \partial{ c_{i} } }{ \partial{y} }  \right) = k_{i}\epsilon^{2}\frac{\partial}{\partial{x}} \left( \frac{\partial{c_{i}}}{\partial{x}} + z_{i}c_{i} \frac{\partial{\phi}}{\partial{x}}\right) + k_{i} \frac{\partial}{\partial{y}} \left( \frac{\partial{c_{i}}}{\partial{y}} + z_{i}c_{i} \frac{\partial{\phi}}{\partial{y}}\right) ,\label{17}
	\end{align}
	with initial condition  
	$c_{i}\left(x,y,0\right) = c_{I,i}(x)$ and the boundary condition 
	\begin{align}
		\left( \mp \epsilon \beta \hat{i} + \hat{j} \right) \cdot \left\{ \epsilon \left( \frac{\partial{c_{i}}}{\partial{x}} + z_{i}c_{i} \frac{\partial{\phi}}{\partial{x}}\right)\hat{i} + \left( \frac{\partial{c_{i}}}{\partial{y}} + z_{i}c_{i} \frac{\partial{\phi}}{\partial{y}} \right)\hat{j} \right\}\Bigg|_{\mathbb{R}\times\partial{\Omega}} = 0,
		\hspace{0.3cm} i=1,2,...,N-1 \label{18}
	\end{align}
	and 
	\begin{align}
		\nabla \phi = \left(\frac{\sum\limits_{j=1}^{N-1}(k_{N}-k_{j})z_{j}\nabla c_{j}}{\sum\limits_{j=1}^{N-1}(z_{j}k_{j}-z_{N}k_{N})z_{j}c_{j}}\right), \label{19}
	\end{align}
	where $Pe=\frac{a_{0} U_{0}}{\bar{k}}$ is the Peclet number.
	\section{Multiple-scale perturbation}\label{sec3}
	In this section, a multiple-scale perturbation method is used for the governing equation \eqref{17} and boundary conditions \eqref{18} utilizing the three distinct timescales which are arising from the transport process namely short to long $(i)$ the diffusion time across the channel height $a(x)$, $(ii)$ the advection time across the axial length scale, $(iii)$ the diffusion time across the axial length scale. For three different time scales,
	\begin{align}
		t_{0}=t,\;\;\;\;\; t_{1}=\epsilon t, \;\;\;\;t_{2}=\epsilon^{2}t,\label{20}
	\end{align}
	we can get the fast, medium, and slow time variables, respectively. Then by chain rule, the time derivative can be expanded asymptotically as 
	$$\frac{\partial}{\partial t} \rightarrow\frac{\partial}{\partial t_{0}} + \epsilon \frac{\partial}{\partial t_{1}} + \epsilon^{2} \frac{\partial}{\partial t_{2}},$$ Using the new  differential operator in time, equation \eqref{17} becomes
	\begin{align}
		\frac{ \partial{c_{i}}}{ \partial t_{0} } + 
		\epsilon \frac{ \partial{c_{i}}}{ \partial t_{1} } + \epsilon^{2}\frac{ \partial{c_{i}}}{ \partial t_{2} } +  Pe\;  \epsilon  \left(  u \frac{ \partial{ c_{i} } }{ \partial{x} } + v \frac{ \partial{ c_{i} } }{ \partial{y} }  \right) = k_{i} \epsilon^{2} \frac{\partial}{\partial{x}} \left( \frac{\partial{ c_{i} } }{\partial{x}} + z_{i}c_{i} \frac{\partial{\phi}}{ \partial{x}}\right)\nonumber\\
		+ k_{i} \frac{\partial}{\partial{y}} \left( \frac{\partial{c_{i}}}{\partial{y}} + z_{i}c_{i} \frac{\partial{\phi}}{\partial{y}}\right) \label{21}
	\end{align}
	The shear dispersion problem described by equation (\ref{21}) is solved in the limit $\epsilon <<1$. For the limit, $\epsilon \rightarrow 0$ the asymptotic expansion of $c_{i}$ is 
	\begin{align}
		c_{i}(x,y,t) = c_{i,0}(x, y, t_{0}, t_{1}, t_{2}) + \epsilon\:c_{i,1}(x, y, t_{0}, t_{1}, t_{2}) + \epsilon^{2}\; c_{i,2}(x, y, t_{0}, t_{1}, t_{2})+ O(\epsilon^{3}).\label{22}
	\end{align}
	Substituting the asymptotic expansion of $c_{i}$ into the formula $\nabla\phi$ in equation (\ref{19}) and using the asymptotic expansion of $\phi$,
	\begin{align}
		\phi = \phi_{0} + \epsilon\:\phi_{1} + \epsilon^{2}\; \phi_{2}+ O(\epsilon^{3}). \label{23}
	\end{align}
	In particular, the gradient of the first two coefficients are given by
	\begin{align}   
		\nabla \phi_{0} = \left(\frac{\sum\limits_{j=1}^{N-1}(k_{n}-k_{j})z_{j}\nabla c_{j,0}}{\sum\limits_{j=1}^{N-1}(z_{j}k_{j}-z_{n}k_{n})z_{j}c_{j,0}}\right), \label{24}
	\end{align}
	\begin{align}
		\nabla \phi_{1} = \left(\frac{\sum\limits_{j=1}^{N-1}(k_{n}-k_{j})z_{j}\nabla c_{j,1}}{\sum\limits_{j=1}^{N-1}(z_{j}k_{j}-z_{n}k_{n})z_{j}c_{j,0}}\right) -  \left(\frac{\left(\sum\limits_{j=1}^{N-1}(k_{n}-k_{j})z_{j}\nabla c_{j,0}\right)\left(\sum\limits_{j=1}^{N-1}(k_{n}-k_{j})z_{j} c_{j,1}\right)}{\left(\sum\limits_{j=1}^{N-1}(z_{j}k_{j}-z_{n}k_{n})z_{j}c_{j,0}\right)^{2}}\right),\label{25}
	\end{align}
	Substituting the expansion of $c_{i}$ and $ \nabla\phi$ into the equation \eqref{21} yields an equation in terms of power series of $\epsilon$. Since the equation holds for every small $\epsilon$, the coefficient of each power of $\epsilon$ should be zero. We have performed an order analysis in the preceding systems to analyze the multispecies diffusion.
	\\
	\subsection{Leading order perturbation}
	Considering all the terms of order $O(1)$, and setting the coefficient to be zero, we obtain 
	\begin{align}
		\frac{ \partial{c_{i,0}}}{ \partial{t_{0}} }  =  k_{i} \frac{\partial}{\partial{y}} \left( \frac{\partial{c_{i,0}}}{\partial{y}} + z_{i}c_{i,0} \frac{\partial{\phi_{0}}}{\partial{y}}\right) \label{26}
	\end{align}
	With the initial condition
	\begin{align}
		c_{i,0}\left(x, y, 0, 0, 0\right)=c_{I,i}(x) \label{27}
	\end{align}
	and boundary condition
	\begin{align}
		k_{i} \frac{\partial}{\partial{y}} \left( \frac{\partial{c_{i,0}}}{\partial{y}} + z_{i}c_{i,0} \frac{\partial{\phi_{0}}}{\partial{y}}\right)\Bigg|_{y=\pm a(x)} =0. \label{28}
	\end{align}
	The initial condition is a function of the variable $x$ only and the fast process, that is, in short time (diffusion in $y$ direction) equilibrates quickly as compared to the slow process. Owing of this equations \eqref{26} - \eqref{28} indicate that leading order concentrations are independent of $y$ and $t_{0}$, that is
	\begin{align}
		c_{i,0}(x, y, t_{0}, t_{1}, t_{2})=c_{i,0}(x,t_{1},t_{2}),\;\;\; i=1,2,...,N. \label{29}
	\end{align}
	Physically, the independence of $c_{i, 0}$'s on the transverse coordinate y implies that $c_{i, 0}$'s are representing the transverse averaged concentrations.
	\subsection{First order perturbation}
	Considering all the terms of order $O(\epsilon^{1})$, that is, in the medium time scale, we obtained
	\begin{align}
		\frac{ \partial{c_{i,1}}}{ \partial t_{0} } + \frac{ \partial{c_{i,0}}}{ \partial t_{1} } +    Pe   \left(  u \frac{ \partial{ c_{i,0} } }{ \partial{x} } + v \frac{ \partial{ c_{i,0} } }{ \partial{y} }  \right) = k_{i}   \frac{\partial}{\partial{y}} \left( \frac{\partial{c_{i,1}}}{\partial{y}} + z_{i}c_{i,0} \frac{\partial{\phi_{1}}}{\partial{y}} + z_{i}c_{i,1} \frac{\partial{\phi_{0}}}{\partial{y}}\right), \label{30}
	\end{align} 
	with initial $c_{i,1}\left(x, y, 0, 0, 0\right)=0$ and the no flux boundary condition
	\begin{align}
		k_{i}\left( \frac{\partial{c_{i,1}}}{\partial{y}} + z_{i}c_{i,1} \frac{\partial{\phi_{0}}}{\partial{y}} + z_{i}c_{i,0} \frac{\partial{\phi_{1}}}{\partial{y}} \right)\Bigg|_{y=\pm a(x)} = 0,
		\hspace{0.3cm} i=1,2,...,N-1.\label{31}
	\end{align}
	Since $c_{i,0}$ is independent of $y$ and $t_{0}$, then from the equation \eqref{24} and \eqref{25} we get
	\begin{align}   
		\frac{\partial \phi_{0}}{\partial y} = 0, \hspace{0.5cm} \frac{\partial \phi_{1}}{\partial y} = \left(\frac{\sum\limits_{j=1}^{N-1}(k_{n}-k_{j})z_{j} \frac{\partial c_{j,1}}{\partial y}}{\sum\limits_{j=1}^{N-1}(z_{j}k_{j}-z_{n}k_{n})z_{j}c_{j,0}}\right).\label{32}
	\end{align}
	Therefore equation \eqref{30} becomes
	\begin{align}
		\frac{ \partial{c_{i,1}}}{ \partial t_{0} } + \frac{ \partial{c_{i,0}}}{ \partial t_{1} } +   Pe\:u \frac{ \partial{ c_{i,0} } }{ \partial{x} }  = k_{i}   \frac{\partial}{\partial{y}} \left( \frac{\partial{c_{i,1}}}{\partial{y}} + z_{i}c_{i,0} \frac{\partial{\phi_{1}}}{\partial{y}}\right),\hspace{0.3cm} i=1,2,...,N-1, \label{33}
	\end{align}
	Now, let us define the average over a period of time 
	$t_{0}$ 
	\begin{align}
		\overline{(\cdot)} = \frac{1}{t_{0}}\int_{0}^{t_{0}}(\cdot) dt_{0} \;, \label{34}
	\end{align}
	and the transverse averaging as 
	\begin{align}
		\left<(\cdot)\right> = \frac{1}{2a(x)}\int_{-a(x)}^{a(x)}(\cdot) dy \label{35}
	\end{align}
	Taking the time average of the equations \eqref{33} and \eqref{31},
	\begin{align}
		\frac{ \partial{c_{i,0}}}{ \partial t_{1} } +   Pe\:\overline{u} \frac{ \partial{ c_{i,0} } }{ \partial{x} }  = k_{i}   \frac{\partial}{\partial{y}} \left( \frac{\partial \overline{c_{i,1}}}{\partial{y}} + z_{i}c_{i,0} \frac{\partial{\overline\phi_{1}}}{\partial{y}}\right),\label{36}
	\end{align}
	\begin{align}
		k_{i}\left( \frac{\partial\overline{c_{i,1}}}{\partial{y}} + z_{i}c_{i,0} \frac{\partial\overline{\phi_{1}}}{\partial{y}} \right)\Bigg|_{y=\pm a(x)} =\;\;\; 0,
		\hspace{0.3cm} i=1,2,...,N-1. \label{37}
	\end{align}
	Further considering the transverse average of \eqref{36} and we get after using \eqref{37}
	\begin{align}
		\frac{ \partial{c_{i,0}}}{ \partial t_{1} } +   Pe\: \left<\overline{u}\right> \frac{ \partial{ c_{i,0} } }{ \partial{x} } =0,\label{38}
	\end{align}
	To obtain the solution for the first order concentrations $c_{i,1}$, we substitute equation \eqref{38} in equation \eqref{33},
	\begin{align}
		\frac{ \partial{c_{i,1}}}{ \partial t_{0} } + Pe\:  \left( u-\left<\overline{u}\right> \right) \frac{ \partial{ c_{i,0} } }{ \partial{x} } =  k_{i}   \frac{\partial}{\partial{y}} \left( \frac{\partial{c_{i,1}}}{\partial{y}} + z_{i}c_{i,0} \frac{\partial{\phi_{1}}}{\partial{y}}\right),\hspace{0.3cm} i=1,2,...,N-1, \label{39}
	\end{align}
	which is a linear equation of $c_{i,1}$ and can be written as\\
	$$ \frac{\partial \mathbf{c_{1}}}{\partial t_{0} } + Pe\; \left( u-\left<\overline{u}\right> \right)  \;\frac{\partial\mathbf{c_{0}}}{\partial x} = \mathbf{D}\Delta_{y}\mathbf{c_{1}},$$
	or
	\begin{align}
		(-\frac{\partial}{\partial t_{0} } +\mathbf{D}\Delta_{y})\mathbf{c_{1}} =  Pe\; \left( u-\left<\overline{u}\right> \right) \;\frac{\partial\mathbf{c_{0}}}{\partial x} \label{40}
	\end{align}
	In this equation the cross-stream variation can be represented by $c_{i,1}$ which is governed by the differential streamwise advection of the leading-order concentrations relative to the mean velocity $ \left<\overline{u}\right>$.
	\begin{align}
		\mathbf{D} =
		\begin{bmatrix}
			k_{1} & 0 & ... & 0 \\
			0 & k_{2} & ... & 0 \\
			... & ... & ... & ...\\
			... & ... & ... & ...\\
			0 & 0 & ... &  k_{N-1}              \end{bmatrix}
		-
		\begin{bmatrix}
			k_{1}z_{1}c_{1,0} \\
			k_{2}z_{2}c_{2,0} \\
			........          \\
			...........          \\
			k_{N-1}z_{N-1}c_{N-1,0}           \end{bmatrix}
		\frac{\begin{bmatrix}
				(k_{1}-k_{N})z_{1} & ..... & (k_{N-1}-k_{N})z_{N-1}
		\end{bmatrix}}{\sum\limits_{j=1}^{N-1}(z_{j}k_{j}-z_{N}k_{N})z_{j}c_{j,0}}, \label{41}   
	\end{align}
	where $\mathbf{c_{0}}=(c_{1,0},\;c_{2,0},...,c_{N-1,0})$, $\mathbf{c_{1}}=(c_{1,1},\;c_{2,1},...,c_{N-1,1})$, $ \Delta_{y} = \frac{ \partial^{2} }{ \partial y^{2} } $ and $ \Delta_{y}\mathbf{c_{1}} = \left( \frac{\partial^{2} }{\partial y^{2}}(c_{1,1}),\; \frac{\partial^{2} }{\partial y^{2}}(c_{1,2}),...,\frac{\partial^{2} }{\partial y^{2}}(c_{N-1,1}) \right)$. Hence, $\mathbf{D}$ is the difference between a diagonal matrix and the product of two vectors.\\
	Fredholm's solvability states that the linear equation $L\psi = f$ has a solution if and only if the function $f$ is orthogonal, that is, $\left<f,g\right>=0$ to every solution of the adjoint equation $L^*g=0$, where $L^{*}$ is the adjoint operator of $L$. Here, the constant function solves the adjoint problem, and then we obtain the solvability condition for the equation \eqref{40} as, 
	\begin{align}
		\left< \overline{ Pe\; \left( u-\left<\overline{u}\right> \right)  \;\frac{\partial\mathbf{c_{0}}}{\partial x} }\right > = Pe\; \left< \overline{ \left( u-\left<\overline{u}\right> \right) }  \right > \;\frac{\partial\mathbf{c_{0}}}{\partial x} = 0. \label{42}  
	\end{align}
	Therefore, the solution for the equation \eqref{40} can be expressed as
	\begin{align}
		\mathbf{c_{1}} = Pe \; (-\frac{\partial}{\partial t_{0} } +\mathbf{D}\Delta_{y})^{-1}\left( \tilde{u}\; \frac{\partial\mathbf{c_{0}}}{\partial x} \right), \hspace{0.5cm} \tilde{u} = \left( u-\left<\overline{u}\right> \right) \label{43}
	\end{align}
	In this problem, we consider the steady shear flow, and then from the equation \eqref{43}, we can obtained 
	$$
	\mathbf{c_{1}} = Pe \;  \Delta_{y}^{-1}(\tilde{u}(x,y)) \; \mathbf{D}^{-1}\frac{\partial \mathbf{c_{0}}}{\partial x} 
	$$
	or
	\begin{align}
		\mathbf{c_{1}} = Pe \; A(x,y) \; \mathbf{D}^{-1}\frac{\partial \mathbf{c_{0}}}{\partial x},\hspace{0.2 cm} A(x,y)=\Delta_{y}^{-1}(\tilde{u}(x,y)) . \label{44}
	\end{align}
	The function $A(x,y)$ is representing a boundary value problem, obtained by substituting equation \eqref{44} into equation \eqref{40} for the governing equation and is represented by 
	\begin{align}
		\Delta_{y}A=\frac{\partial ^{2} A}{\partial y^{2}}=\tilde{u}(x,y) . \label{45A}
	\end{align}
	Equation \eqref{44} is substitute in equation \eqref{31}  for the boundary condition as,
	\begin{align}
		\frac{\partial A}{\partial y}\Bigg|_{y=\pm a(x)}=0. \label{45}
	\end{align}
	By using the Sherman-Morrison formula \cite{sherman1950adjustment, akgun2001fast}, the change in the inverse of a matrix $K$ due to a change $\delta K$, which is of rank one, may be written as $\delta K = uv^{T}$. Therefore, the matrix $\mathbf{D}$ can be inverted as,
	\begin{align}
		\mathbf{D}^{-1} =
		\begin{bmatrix}
			\frac{1}{k_{1}} & 0 & ... & 0 \\
			0 & \frac{1}{k_{2}} & ... & 0 \\
			... & ... & ... & ...\\
			... & ... & ... & ...\\
			0 & 0 & ... &  \frac{1}{k_{N-1}}              \end{bmatrix}\nonumber\\
		\times \left( I_{N-1} +
		\begin{bmatrix}
			k_{1}z_{1}c_{1,0} \\
			k_{2}z_{2}c_{2,0} \\
			........          \\
			...........          \\
			k_{N-1}z_{N-1}c_{N-1,0}           \end{bmatrix}
		\frac{ \begin{bmatrix}
				\frac{(k_{1}-k_{N})z_{1}}{k_{1}} & ..... & \frac{(k_{N-1}-k_{N})z_{N-1}}{k_{N-1}}
		\end{bmatrix} }{ k_{N}\sum\limits_{j=1}^{N-1}( z_{j} -z_{N} )z_{j}c_{j,0} } \right),\label{46} 
	\end{align}
	where $I_{N-1}$ is the $(N-1)\times(N-1)$ identity matrix.
	\subsection{Second order perturbation}
	Considering the second order terms of the expansion of the species transport equation \ref{21}, it can be note that, unlike $c_{i,0}(x,t_{1},t_{2})$, the additional dependence of $c_{i,1}(x,y,t_{0},t_{1},t_{2})$ on $y$ and $t_{0}$ indicates that $c_{i,1}$, $c_{i,2}$, and other higher order terms in the expansion of equation \eqref{22} are fluctuating components, that must have a zero transverse average, i.e.,
	\begin{align}
		\left<c_{i,j}\right> = 0, \hspace{0.5cm} j = 1,2,...\label{47}
	\end{align}
	Now, grouping all the terms of order 
	$O(\epsilon^{2})$, which are given by in the long time scale equations \eqref{21} and \eqref{18} provides,
	\begin{align}
		\frac{ \partial{c_{i,0}}}{ \partial t_{2} } + \frac{ \partial{c_{i,1}}}{ \partial t_{1} } + \frac{ \partial{c_{i,2}}}{ \partial t_{0} } +  Pe\;  u \frac{ \partial{ c_{i,1} } }{ \partial{x} } + & Pe\; v \frac{ \partial{ c_{i,1} } }{ \partial{y} } = k_{i} \frac{\partial}{\partial{x}} \left( \frac{\partial{ c_{i,0} } }{\partial{x}} + z_{i}c_{i,0} \frac{\partial{\phi_{0}}}{ \partial{x} }\right)\nonumber\\
		& + k_{i} \frac{\partial}{\partial{y}} \left( \frac{\partial{c_{i,2}}}{\partial{y}} + z_{i}c_{i,0} \frac{\partial{\phi_{2}}}{\partial{y}} + z_{i}c_{i,1} \frac{\partial{\phi_{1}}}{\partial{y}} \right) \label{48}
	\end{align}  
	with the boundary condition
	\begin{align}
		k_{i}\left( \frac{\partial{c_{i,2}}}{\partial{y}} + z_{i}c_{i,0} \frac{\partial{\phi_{2}}}{\partial{y}} + z_{i}c_{i,1} \frac{\partial{\phi_{1}}}{\partial{y}} \right)\Bigg|_{y=\pm a(x)} =\;\;\; \pm \beta k_{i}\left( \frac{\partial{ c_{i,0} } }{\partial{x}} + z_{i}c_{i,0} \frac{\partial{\phi_{0}}}{ \partial{x} }\right) \label{49}
	\end{align}
	First, let us consider the fourth term on the left-hand side of equation \eqref{48}. Substituting equation \eqref{44} and taking the time average leads to,
	\begin{align}
		Pe\left(\overline{u \frac{ \partial{ \mathbf{c_{1} } } }{ \partial{x} } } \right) = Pe^{2}\Big[  \overline{u A_{x} } \left( \mathbf{D^{-1}\frac{\partial \mathbf{c_{o}}}{\partial x}} \right) +  \overline{u A } \frac{\partial}{\partial x} \left( \mathbf{D^{-1}} \frac{\partial \mathbf{c_{o}}}{\partial x} \right) \Bigg].\label{50}
	\end{align}
	Further, taking a transverse average gives
	\begin{align}
		Pe\left< \overline{u \frac{ \partial{ \mathbf{c_{1} } } }{ \partial{x} } } \right> = Pe^{2} \Bigg[ \left< \overline{u A_{x} } \right> \left( \mathbf{D^{-1}\frac{\partial \mathbf{c_{o}}}{\partial x}} \right) + \left< \overline{u A } \right> \frac{\partial}{\partial x} \left( \mathbf{D^{-1}} \frac{\partial \mathbf{c_{o}}}{\partial x} \right) \Bigg].\label{51}
	\end{align}
	Similarly, time and transverse averaging of the fifth term on the left-hand side of equation \eqref{48} gives
	\begin{align}
		Pe\left< \overline{v \frac{ \partial{ \mathbf{c_{1} } } }{ \partial{y} } } \right> = Pe^{2} \Bigg[ \left< \overline{v A_{y} }  \right> \left( \mathbf{D^{-1}\frac{\partial \mathbf{c_{o}}}{\partial x}} \right) \Bigg],\label{52}
	\end{align}
	where the subscript $x$ and $y$ denotes the first order derivative with respect to $x$ and $y$, respectively.\\
	After taking the time average equation \eqref{48} and \eqref{49} becomes
	\begin{align} 
		\frac{ \partial{ \overline{c_{i,2}}}}{ \partial t_{0} } + \frac{ \partial{ \overline{c_{i,1}}} }{ \partial t_{1} } +  \frac{ \partial{c_{i,0}}}{ \partial t_{2} } +  Pe\; & \left(  \overline{u \frac{ \partial{ c_{i,1}  } }{ \partial{x} } } + \overline{v \frac{ \partial{ c_{i,1} } }{ \partial{y} } }  \right)  = k_{i} \frac{\partial}{\partial{x}} \left( \frac{\partial{ c_{i,0} } }{\partial{x}} + z_{i}c_{i,0} \frac{\partial{\phi_{0}}}{ \partial{x} }\right)\nonumber\\
		& +\; k_{i} \frac{\partial}{\partial{y}} \left( \frac{\partial{\overline{c_{i,2}}}}{\partial{y}} + z_{i}c_{i,0} \frac{\partial{\overline{\phi_{2}}}}{\partial{y}} + z_{i}\;\overline{c_{i,1} \frac{\partial{\phi_{1}}}{\partial{y}}} \right)  \label{53}
	\end{align}
	\begin{align}
		k_{i}\left( \frac{\overline{\partial{c_{i,2}}}}{\partial{y}} + z_{i}c_{i,0} \frac{\partial{\overline{\phi_{2}}}}{\partial{y}} + z_{i}\; \overline{c_{i,1} \frac{\partial{\phi_{1}}}{\partial{y}}} \right)\Bigg|_{y=\pm a(x)} =\;\;\; \pm \beta k_{i}\left( \frac{\partial{ c_{i,0} } }{\partial{x}} + z_{i}c_{i,0} \frac{\partial{\phi_{0}}}{ \partial{x} }\right) \label{54}
	\end{align}
	Moreover, considering the transverse average of equation \eqref{53} and using equation \eqref{47}, the first and second terms on the left-hand side of equation \eqref{53} become zero. Furthermore, by applying the boundary condition given in equation \eqref{54}, the second term on the right-hand side of equation \eqref{53} combines with the first term in the right-hand side of equation \eqref{53}, we obtained
	\begin{align}
		\frac{ \partial{c_{i,0}}}{ \partial t_{2} } + Pe\;  \left(  \left< \overline{u \frac{ \partial{ c_{i,1}  } }{ \partial{x} } } \right> + \left< \overline{v \frac{ \partial{ c_{i,1} } }{ \partial{y} } } \right>  \right)
		= \left( 1 + \frac{\beta}{a(x)}\right)k_{i} \frac{\partial}{\partial{x}} \left( \frac{\partial{ c_{i,0} } }{\partial{x}} + z_{i}c_{i,0} \frac{\partial{\phi_{0}}}{ \partial{x} }\right),\nonumber
	\end{align}
	for $ i = 1,2,3...,N-1.$          
	\begin{align}
		\frac{ \partial \mathbf{{c_{0}}}}{ \partial t_{2} } + Pe\;  \left(  \left< \overline{u \frac{ \partial \mathbf{ { c_{1} }  } }{ \partial{x} } } \right> + \left< \overline{v \frac{ \partial{ \mathbf{ c_{1} } } }{ \partial{y} } } \right>  \right) = \left( 1 + \frac{\beta}{a(x)}\right) \frac{\partial}{\partial{x}} \left( \mathbf{D} \frac{\partial \mathbf{{ c_{0} }} }{\partial{x}}\right). \label{55}
	\end{align}
	Substituting equation \eqref{51} and \eqref{52} into the equation \eqref{55} we get
	\begin{align}
		\frac{ \partial \mathbf{{c_{0}}}}{ \partial t_{2} } + Pe^{2} \Bigg[ \left( \left< \overline{u A_{x} } \right> + \left< \overline{v A_{y} } \right> \right) \left( \mathbf{D^{-1}\frac{\partial \mathbf{c_{o}}}{\partial x}} \right) + \left< \overline{u A } \right> \frac{\partial}{\partial x} \left( \mathbf{D^{-1}} \frac{\partial \mathbf{c_{o}}}{\partial x} \right)  \Bigg]\nonumber\\   
		= \left( 1 + \frac{\beta}{a(x)}\right) \frac{\partial}{\partial{x}} \left( \mathbf{D} \frac{\partial \mathbf{{ c_{0} }} }{\partial{x}}\right).\label{56}
	\end{align}
	After simplifying the above equation, we get 
	\begin{align}
		\frac{ \partial \mathbf{{c_{0}}}}{ \partial t_{2} } +  \Bigg[ Pe^{2} \left( \left< \overline{u A_{x} } \right> \mathbf{ D^{-1} } + \left< \overline{v A_{y} } \right>  \mathbf{ D^{-1} }   + \left< \overline{u A } \right> \frac{\partial \mathbf{ D^{-1} } }{\partial {x} } \right) - \left( 1 + \frac{\beta}{a(x)}\right) \frac{\partial \mathbf{ D } }{\partial {x} }    \Bigg] \frac{\partial \mathbf{c_{o}}}{\partial x}\nonumber\\  
		= \Bigg[ \left( 1 + \frac{\beta}{a(x)} \right) \mathbf{D} - Pe^{2}\left< \overline{u A } \right>  \mathbf{ D^{-1} }  \Bigg]  \frac{ \partial^{2} \mathbf{c_{o}} }{ \partial{x^{2}} }.\label{57}
	\end{align}
	Equation \eqref{57} represent the generalized averaged transport system of equations for multispecies electrolyte solutions which is the key result of the present paper. The domain of the computation for the concentration distribution is considered within the arbitrarily shaped axisymmetric channel and the flow fields are assumed to very on the streamwise ($x$) and the transverse ($y$) coordinate .\\
	The non-linear equation \eqref{57} is an approximation of \eqref{3} within the 
	limit $\epsilon<<1$, as well as at long times.
	\\
	For further simplification, we employ the inversion of the Laplace operator $\Delta_{y}^{-1}(\tilde{u})$ which depends on the channel geometry. In this case, the channel domain chosen as, $\Omega=\{y:y\in[-a(x), a(x)]\}$ and hence 
	\begin{align}
		A= \Delta_{y}^{-1}({\tilde{u})} = \int_{-a(x)}^{y}\int_{-a(x)}^{s_{2}}\tilde{u}(x,s_{1})ds_{1}ds_{2} \;+ f(x).\label{58}
	\end{align}
	To determine $f(x)$, we need an additional condition for which a normalization condition is imposed as,
	\begin{align}
		\int_{-a(x)}^{a(x)}A(x,y)dy=0.\label{59}
	\end{align}
	Now using the equation \eqref{14} into equation \eqref{58} yields,
	\begin{align}
		\tilde{u}=\frac{Q}{4a(x)}\left( 1-\frac{3y^{2}}{a(x)^{2}} \right),\;\;\; A(x,y)=\frac{Q}{4}\left(  \frac{y^{2}}{2a(x)} - \frac{y^{4}}{4a(x)^{3}} - \frac{7a(x)}{60}  \right). \label{60}
	\end{align}
	The concentration evolution can only be solved after estimating the averaged terms in Eq.\eqref{57}. The required terms are
	\begin{align}
		\left< \overline{u A_{x} } \right> = -\frac{2Q^{2}\beta}{105a(x)},\hspace{0.5cm} \left< \overline{u A } \right> = -\frac{Q^{2}\beta}{210}, \hspace{0.5cm} \left< \overline{v A_{y} } \right> = \frac{3Q^{2}\beta}{210a(x)}.\label{61}
	\end{align}
	Therefore after using \eqref{61} into the equation \eqref{57}, becomes 
	\begin{align}
		\frac{ \partial \mathbf{{c_{0}}}}{ \partial t_{2} } - Pe^{2}\Bigg[ \frac{Q^{2}\beta}{210a(x)} \left(   \mathbf{D^{-1}} \frac{\partial \mathbf{c_{0}} } {\partial x}  \right) + \frac{Q^{2}}{210}  \frac{\partial  }{\partial {x} } \left(  \mathbf{ D^{-1} } \frac{\partial \mathbf{c_{o}}}{\partial x} \right)   \Bigg]  \nonumber\\
		= \left( 1 + \frac{\beta}{a(x)} \right)   \frac{\partial  }{\partial {x} } \left(\mathbf{ D } \frac{\partial \mathbf{c_{o}}}{\partial x} \right) \label{62}      
	\end{align}
	or
	\begin{align}
		\frac{ \partial \mathbf{{c_{0}}}}{ \partial t_{2} } - \Bigg[ \frac{Pe^{2}Q^{2}}{210} \left( \frac{\beta} {a(x)}  \mathbf{D^{-1}} + \frac{\partial } {\partial x}\mathbf{D^{-1}}  \right) + \left( 1 + \frac{\beta}{a(x)}\right) \frac{\partial  }{\partial {x} }  \mathbf{ D }  \Bigg] \frac{\partial \mathbf{c_{0}}}{\partial x} \nonumber\\  
		= \Bigg[ \left( 1 + \frac{\beta}{a(x)} \right) \mathbf{D} + \frac{Pe^{2}Q^{2}}{210} \mathbf{D^{-1}} \Bigg]  \frac{ \partial^{2} \mathbf{c_{0}} }{ \partial{x^{2}} }. \label{63}
	\end{align}
	Eq. \eqref{63} is the simplified version of the full Nernst-Planck equation \eqref{7}, used to describe the average concentrations of $N$ ionic species change over time due to the combined effects of Poiseuille flow advection, and ionic interactions. The reduced order model for this equation is obtained through a time scale analysis conducted and the highest order equation that is considered here is equation \eqref{48} of $O(\epsilon^2)$, but it can be observed that the reduced order Eq. \eqref{63} is correct up to $O(\epsilon^{3})$.\\
	Many physical systems, particularly in electrochemistry and microfluidics, involve three ionic species, such as ternary electrolyte solutions or mixtures of two binary electrolytes i,e. the mixture of sodium chloride and sodium fluorescein. In these case where $N=3$, the diffusion tensor $\mathbf{D}$ mentioned in equation \eqref{41}, along with its inverse, becomes explicitly dependent on the local ionic concentrations and can be represented by is of the form
	\begin{align} 
		\mathbf{D} =
		\begin{bmatrix}
			k_{1}-\frac{c_{1}k_{1}(k_{1}-k_{3})z_{1}^{2}}{c_{1}z_{1}(k_{1}z_{1}-k_{3}z_{3}) + c_{2}z_{2}(k_{2}z_{2}-k_{3}z_{3})}    &    -\frac{c_{1}k_{1}(k_{2}-k_{3})z_{1}z_{2}}{c_{1}z_{1}(k_{1}z_{1}-k_{3}z_{3}) + c_{2}z_{2}(k_{2}z_{2}-k_{3}z_{3})}  \\
			\\ -\frac{c_{2}k_{2}(k_{1}-k_{3})z_{1}z_{2}}{c_{1}z_{1}(k_{1}z_{1}-k_{3}z_{3}) + c_{2}z_{2}(k_{2}z_{2}-k_{3}z_{3})}    &   k_{2}-\frac{c_{2}k_{2}(k_{2}-k_{3})z_{2}^{2}}{c_{1}z_{1}(k_{1}z_{1}-k_{3}z_{3}) + c_{2}z_{2}(k_{2}z_{2}-k_{3}z_{3})}             \end{bmatrix} \nonumber\\
		\text{and}\hspace{0.5cm}
		\mathbf{D^{-1}} =
		\begin{bmatrix}
			\frac{c_{2}k_{3}z_{2}(z_{2}-z_{3})+c_{1}z_{1}(k_{1}z_{1}-k_{3}z_{3})} {k_{1}k_{3}((c_{1}z_{1}(z_{1}-z_{3}) + c_{2}z_{2}(z_{2}-z_{3}))}    &    \frac{c_{1}(k_{2}-k_{3})z_{1}z_{2}} {k_{1}k_{3}((c_{1}z_{1}(z_{1}-z_{3}) + c_{2}z_{2}(z_{2}-z_{3}))}  \\
			\\ \frac{c_{2}(k_{1}-k_{3})z_{1}z_{2}} { k_{1}k_{3}((c_{1}z_{1}(z_{1}-z_{3}) + c_{2}z_{2}(z_{2}-z_{3})) }   &   \frac{c_{1}k_{3}z_{1}(z_{1}-z_{3})+c_{2}z_{2}(k_{2}z_{2}-k_{3}z_{3})} {k_{1}k_{3}((c_{1}z_{1}(z_{1}-z_{3}) + c_{2}z_{2}(z_{2}-z_{3}))}           \end{bmatrix}. \label{64}
	\end{align}
	This concentration-dependent coupling significantly influences the species transport behavior.
	\section{Validation}\label{sec4}
	The governing equation for species transport involves highly nonlinear terms and coupling terms associated with variables. To solve numerically, we take the initial conditions, diffusivities, and valences to be the following form:
	\begin{align}
		c_{I,1}=c_{I,2}=\frac{exp(-\frac{1}{2}(\frac{x}{\sigma})^2)}{\sigma\sqrt{2\pi} },\nonumber \hspace{2.5cm}\\
		\sigma=\frac{1}{4},\hspace{0.3cm} k_{1}=1,\;\;\;k_{2}=0.1,\;\;\;k_{3}=1, \hspace{0.3cm}z_{1}=1,\;\;\;z_{2}=1,\;\;\;z_{3}=-2.\label{65}
	\end{align}	
	\begin{figure}
		\centering
		\includegraphics[width=0.60\linewidth]{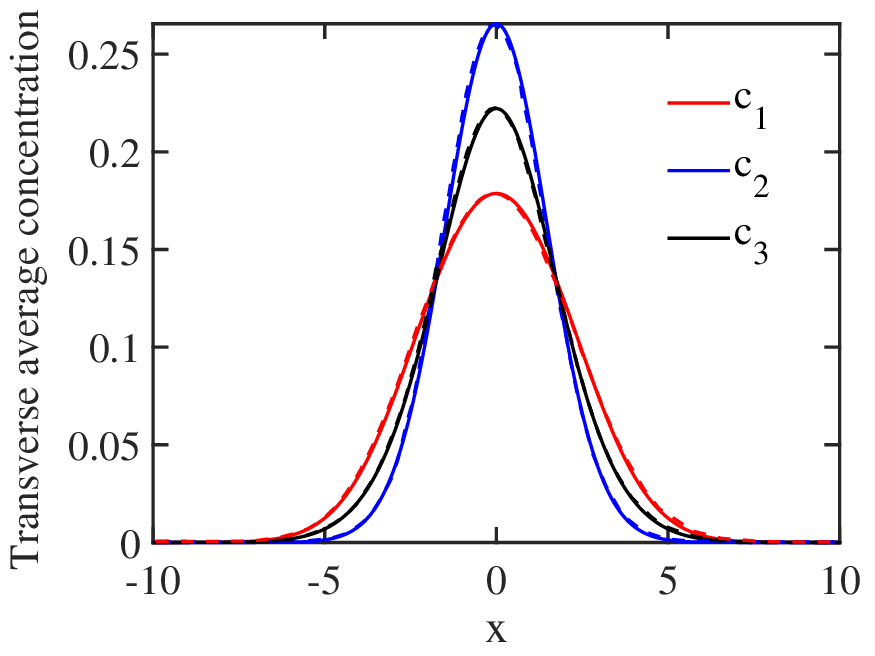}
		\caption{A comparison between the numerical results obtained in the present work (solid lines) and the results published by \cite{ding2023shear} (dashed lines), shown for time t=2.}
		\label{2f}
	\end{figure}
	The mathematical framework given in equation \eqref{63} is validated for the case of a zero-slope channel wall, i.e., for the straight channel. \figurename~\ref{2f} represent a valid comparison between the transverse average concentration profiles from the present study (solid lines) and the benchmark result from \cite{ding2023shear} (dashed lines). The close alignment of the curves confirms the model's accuracy for the baseline geometry.
	\section{ Results and discussions}\label{sec5}
	In this section, the overall outcomes of theoretical frameworks and numerical observations for the species distribution and flow variation indicating the critical range and structural variation. The impact of wall slope variations on the effective dispersion coefficients of multiple charged species in slowly varying, axisymmetric channels, in the absence of an externally applied electric field. The primary theoretical contribution of this study is the derivation of the macrotransport equation \eqref{63}, which describes the coupled transport of three ionic species in a multispecies electrolyte system under electroneutrality and zero-current conditions. A precise understanding of this equation requires careful consideration of its individual components: Includes advection driven by the flow rate $Q$ and novel terms involving the wall slope $\beta=da/dx$, which coupled the geometric variations to both advection and molecular diffusion. The diffusion term includes the nonlinear, concentration-dependent tensor that captures the complex interplay among flow rate, wall slope, and the induced electric potential arising from differences in ionic diffusivities. The resulting non-monotonic effective diffusivities stem from the combined effects of wall-slope variation, differential species diffusion, and electro-geometric coupling are presented. \\
	It is important to emphasise that equation \eqref{63} is valid at the diffusion time scale, once the concentration field becomes homogenised across the channel cross-section; prior to this, the concentration remains non-uniform, and the homogenization result does not hold. For three species, the equation is highly coupled and nonlinear, leading to phenomena absent in binary electrolytes. Numerically, the coupled partial differential equations were solved using the Method of Lines algorithm \cite{schiesser2009compendium}. Finally, the equations are solved in MATLAB by using the inbuilt suite $ode15s$ and setting $reltol$ and $abstol$ as $1.0 \times 10^{-4}$.
	\subsection{Dependence on the wall geometry}
	In this section, the impact of wall geometry on the dispersion mechanism of multispecies electrolyte solutions is discussed. The shape and structure of the confining boundaries influence largely the velocity field within the channel, which in turn control the dispersion of ionic species. Channels with different cross-sectional shapes can lead to distinct dispersion characteristics; and the interaction between the flow, molecular diffusion, and induced electric potential is strongly geometry dependent. Moreover, curved or irregular walls introduce additional complexity, often amplifying or suppressing dispersion as compared to simple straight geometries. These geometric effects highlight the need to carefully account for wall-induced variations when analyzing transport in microfluidic and capillary systems.\\
	In microfluidic systems, capillaries, and lab-on-a-chip devices, the choice of channel shape directly affects efficiency and resolution. By carefully designing or modifying the geometry, dispersion can be minimized to reduce band broadening or improve mixing efficiency, depending on the requirement. It is observed that wall geometry is a key parameter in controlling transport behavior in electrolyte solutions.
	\subsubsection{Sinusoidal channels with different period}
	\begin{figure}
		\centering
		
		\begin{minipage}[t]{0.32\textwidth}
			\centering
			\begin{minipage}[t]{0.1\textwidth}
				\vspace{0pt}
				\textbf{(a)}
			\end{minipage}%
			\begin{minipage}[t]{0.9\textwidth}
				\vspace{0.5cm}
				\includegraphics[width=\linewidth]{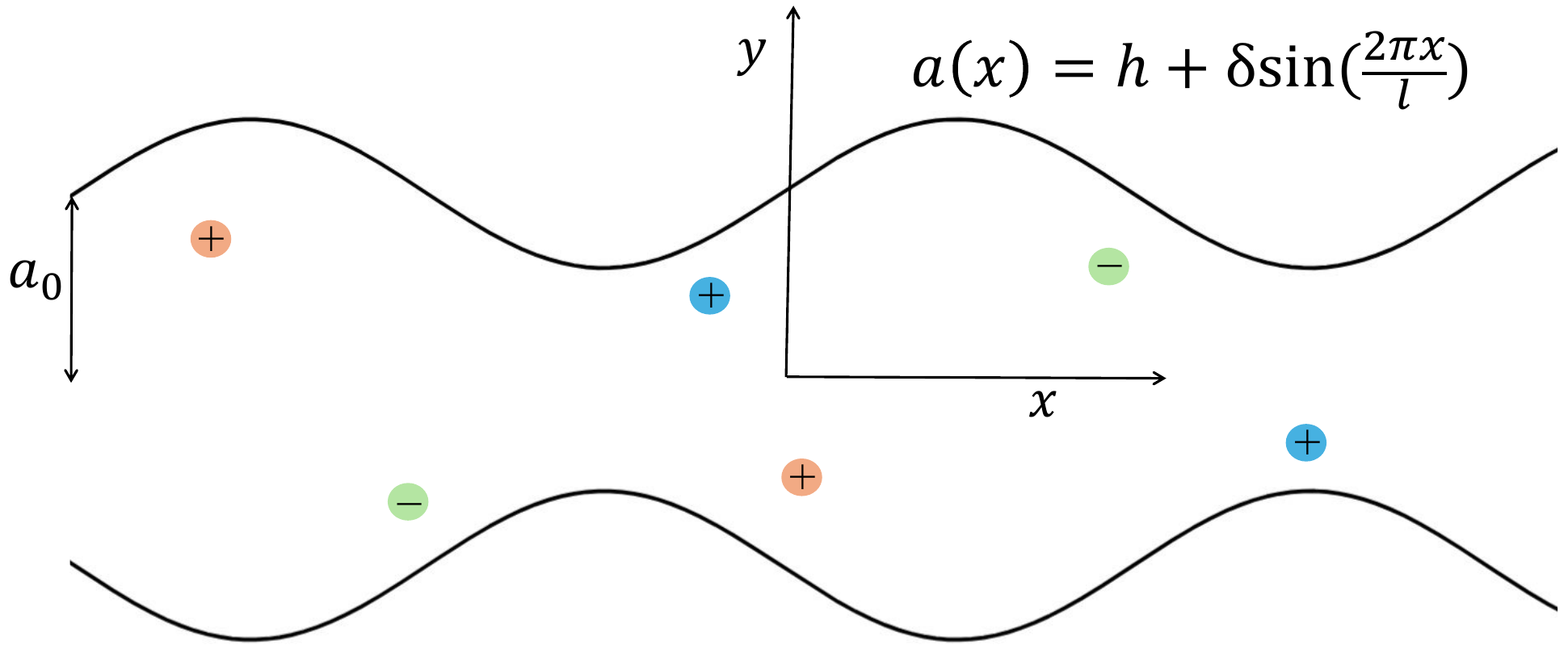}
			\end{minipage}
		\end{minipage}%
		\hfill
		\begin{minipage}[t]{0.32\textwidth}
			\centering
			\begin{minipage}[t]{0.1\textwidth}
				\vspace{0pt}
				\textbf{(b)}
			\end{minipage}%
			\begin{minipage}[t]{0.9\textwidth}
				\vspace{0.5cm}
				\includegraphics[width=\linewidth]{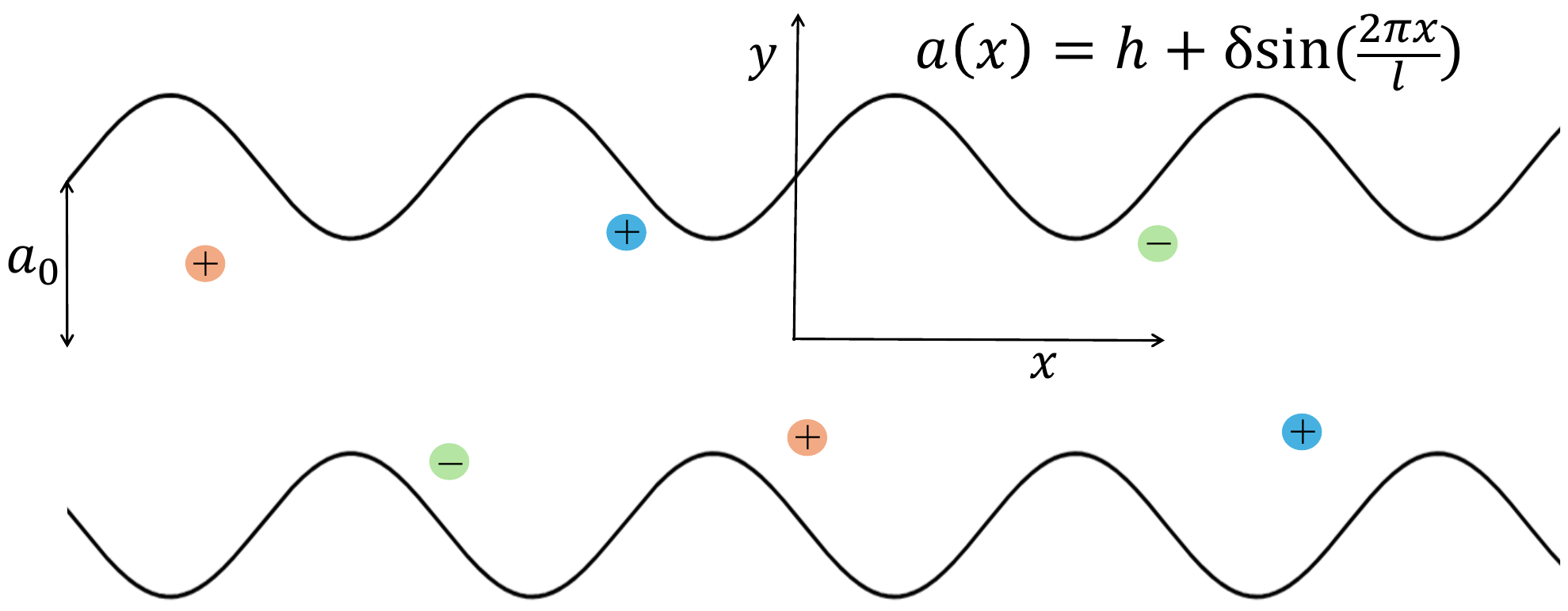}
			\end{minipage}
		\end{minipage}%
		\hfill
		\begin{minipage}[t]{0.32\textwidth}
			\centering
			\begin{minipage}[t]{0.1\textwidth}
				\vspace{0pt}
				\textbf{(c)}  
			\end{minipage}%
			\begin{minipage}[t]{0.9\textwidth}
				\vspace{0.5cm}
				\includegraphics[width=\linewidth]{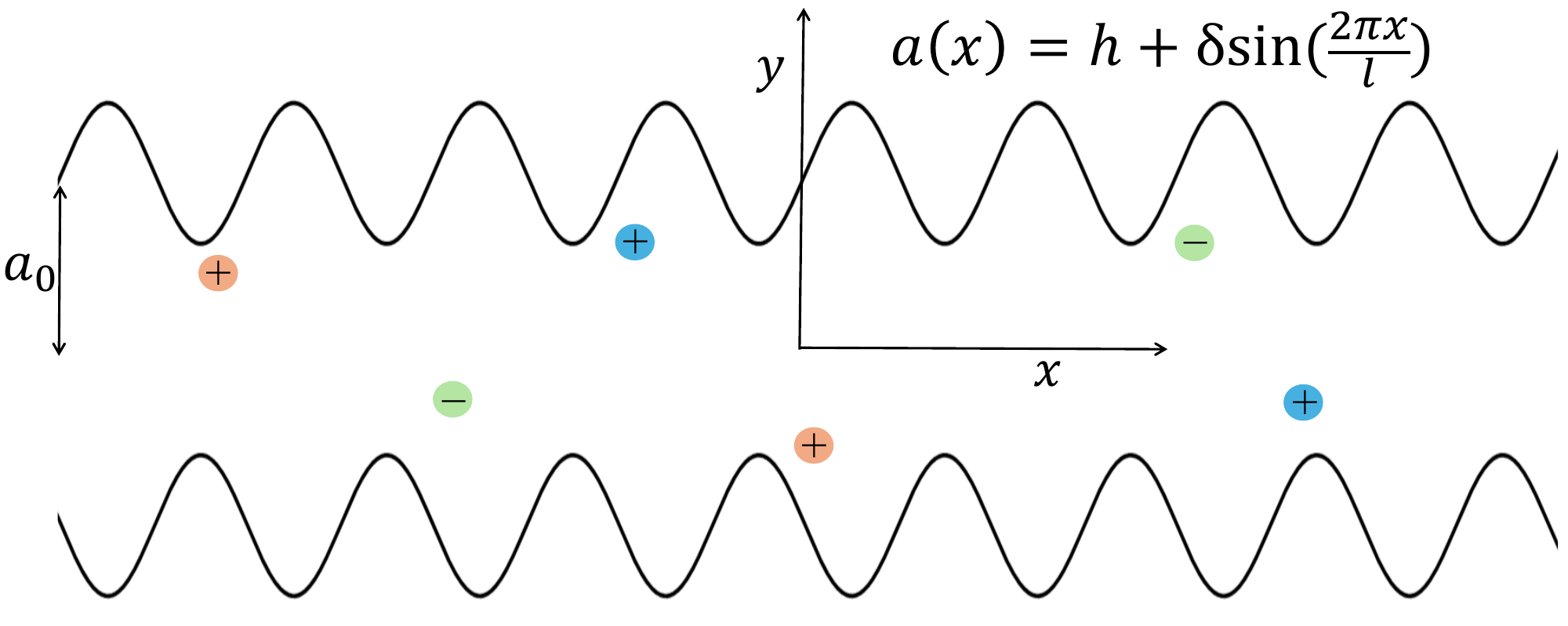}
			\end{minipage}
			\vspace{0.5cm}
		\end{minipage}
		\vspace{0.5cm}
        
		\begin{minipage}[t]{0.32\textwidth}
			\centering
			\begin{minipage}[t]{0.1\textwidth}
				\vspace{0pt}
				\textbf{(d)}
			\end{minipage}%
			\begin{minipage}[t]{0.9\textwidth}
				\vspace{0pt}
				\includegraphics[width=\linewidth]{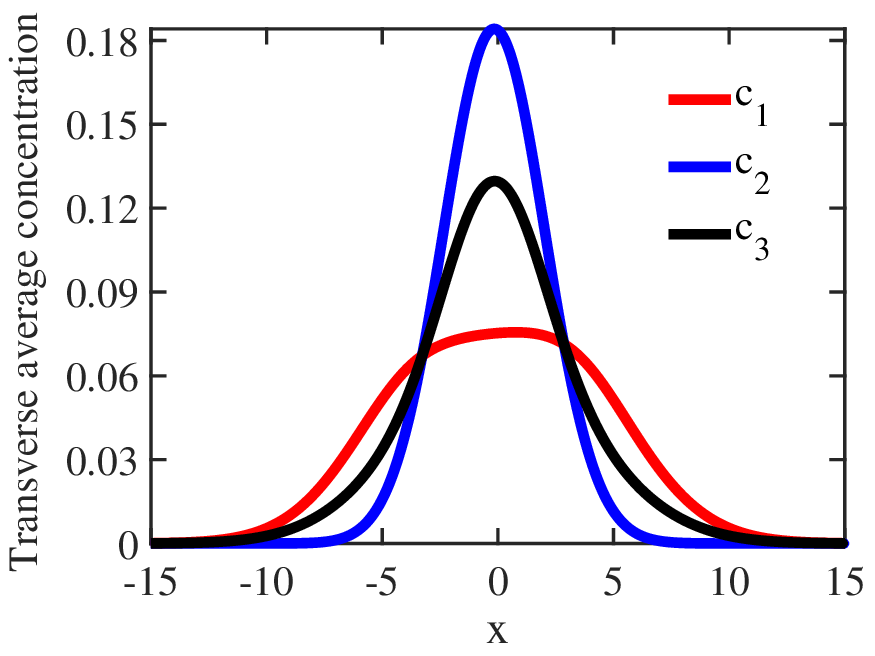}
			\end{minipage}
		\end{minipage}%
		\hfill
		\begin{minipage}[t]{0.32\textwidth}
			\centering
			\begin{minipage}[t]{0.1\textwidth}
				\vspace{0pt}
				\textbf{(e)}
			\end{minipage}%
			\begin{minipage}[t]{0.9\textwidth}
				\vspace{0pt}
				\includegraphics[width=\linewidth]{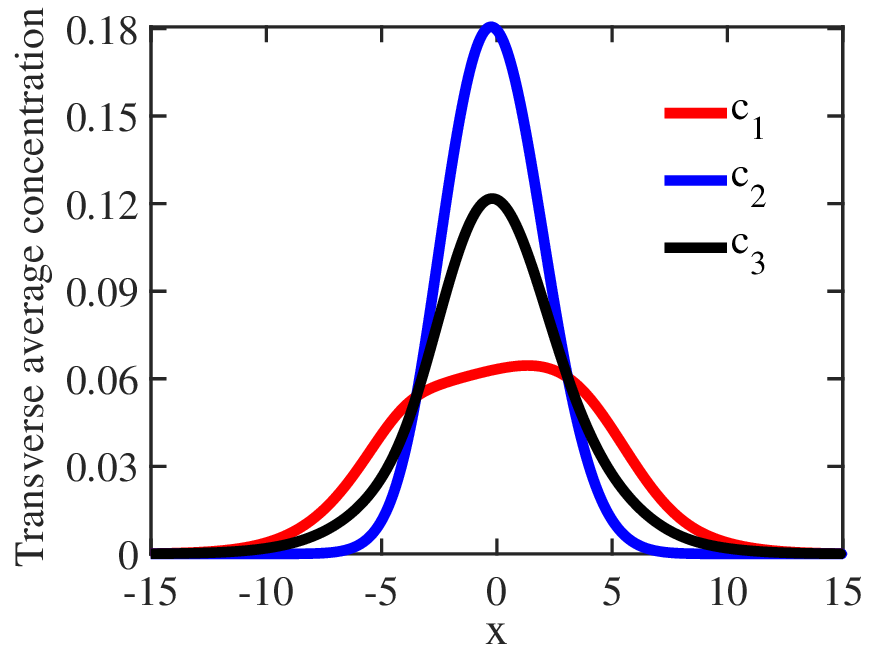}
			\end{minipage}
		\end{minipage}%
		\hfill
		\begin{minipage}[t]{0.32\textwidth}
			\centering
			\begin{minipage}[t]{0.1\textwidth}
				\vspace{0pt}
				\textbf{(f)}  
			\end{minipage}%
			\begin{minipage}[t]{0.9\textwidth}
				\vspace{0pt}
				\includegraphics[width=\linewidth]{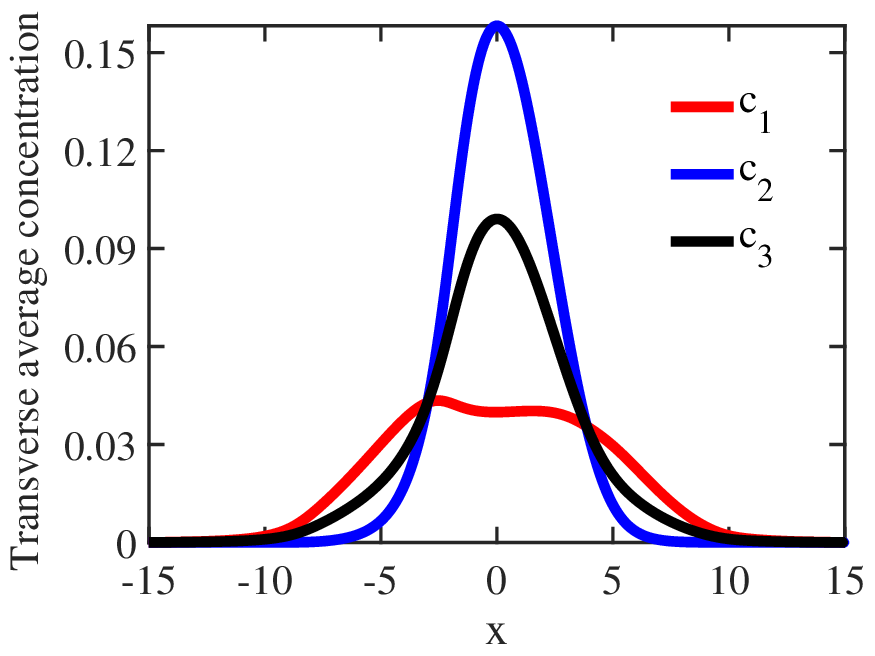}
			\end{minipage}
		\end{minipage}
		\hfill
		\begin{minipage}[t]{0.32\textwidth}
			\centering
			\begin{minipage}[t]{0.1\textwidth}
				\vspace{0pt}
				\textbf{(g)}
			\end{minipage}%
			\begin{minipage}[t]{0.9\textwidth}
				\vspace{0pt}
				\includegraphics[width=\linewidth]{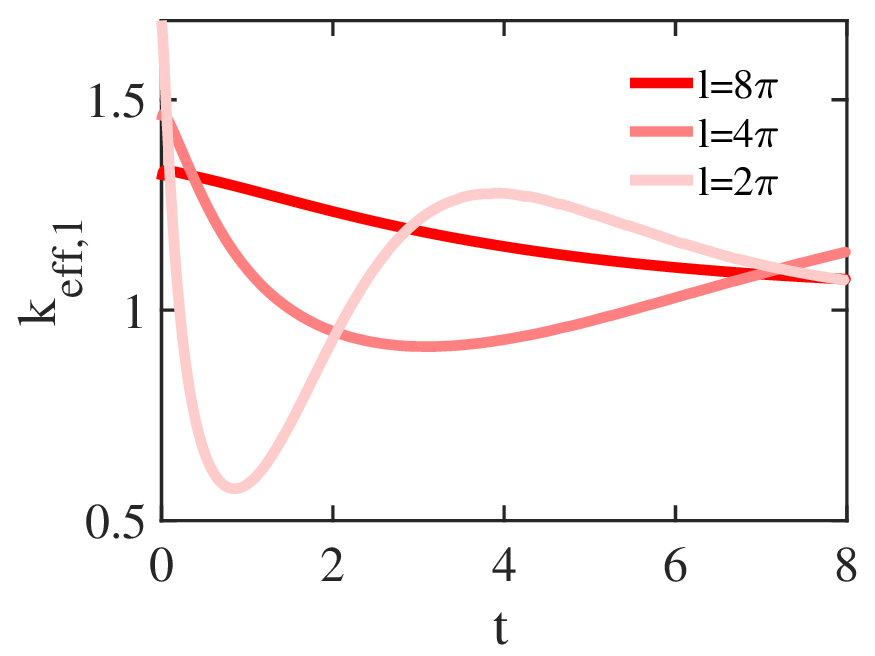}
			\end{minipage}
		\end{minipage}%
		\hfill
		\begin{minipage}[t]{0.32\textwidth}
			\centering
			\begin{minipage}[t]{0.1\textwidth}
				\vspace{0pt}
				\textbf{(h)}
			\end{minipage}%
			\begin{minipage}[t]{0.9\textwidth}
				\vspace{0pt}
				\includegraphics[width=\linewidth]{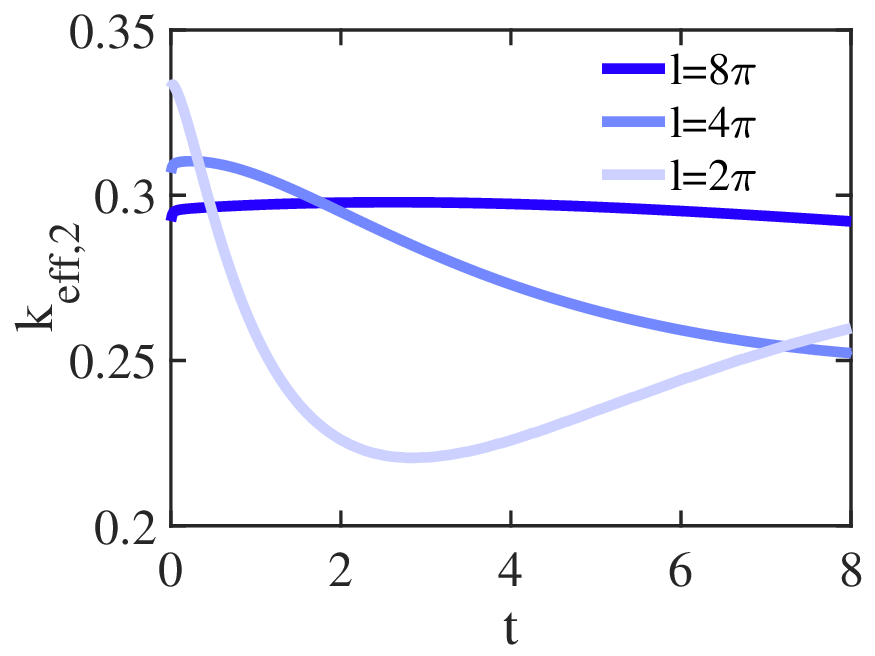}
			\end{minipage}
		\end{minipage}%
		\hfill
		\begin{minipage}[t]{0.32\textwidth}
			\centering
			\begin{minipage}[t]{0.1\textwidth}
				\vspace{0pt}
				\textbf{(i)}  
			\end{minipage}%
			\begin{minipage}[t]{0.9\textwidth}
				\vspace{0pt}
				\includegraphics[width=\linewidth]{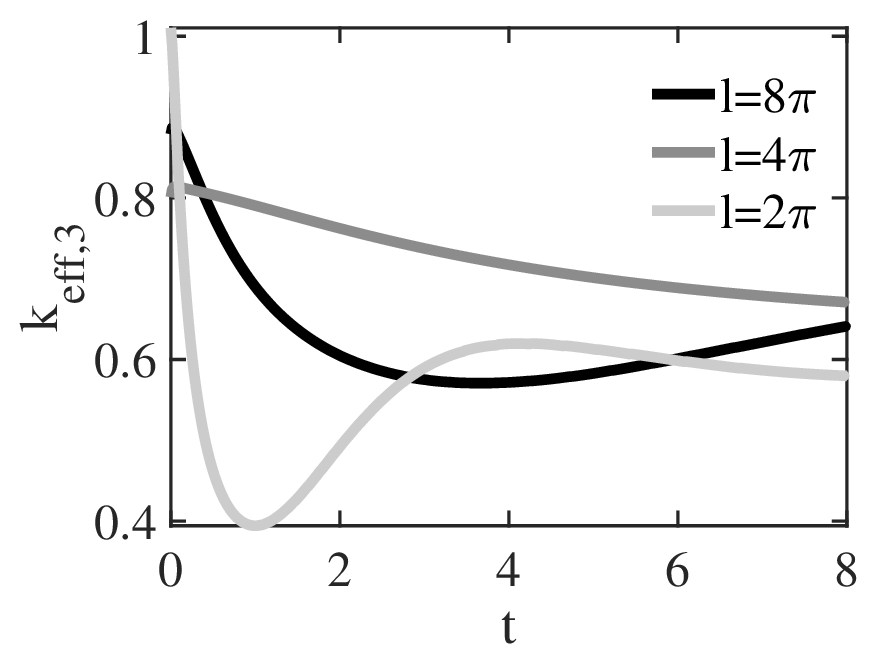}
			\end{minipage}
		\end{minipage}
		\caption{
			Schematic representation of the sinusoidal channel geometry with different periodic lengths, $l$ (a) $l=8\pi$\;(two period), (b) $l=4\pi$\;(four period), and (c) $l=2\pi$\;(eight period), with the inlet height $a_{0}$  and amplitude $\delta=0.5$. (d-e) Represents the transverse-averaged concentration fields along the streamwise direction at time $t=8$ for $Pe=2$ and $Q=1$ corresponding to different periodic lengths. (g-i) Represents temporal evaluation of effective tne diffusivities $k_{eff, i},\;i=1,2,3$  corresponding to the first, second, and third ionic species, respectively, in different periodic lengths. 
		}  
		\label{15f}
	\end{figure}
	The dispersion behavior in sinusoidal channels strongly dependent on the periodic length $l$. For long periods ($l = 8\pi$), geometric variations are mild, resulting in weak modulation of the axial flow. From  \figurename~\ref{15f}, it can be predicted that dispersion is primarily diffusion-dominated, with minimal electric coupling between species. Consequently, species 2 remains sharply localized, species 1 spreads moderately, and effective diffusivities decay smoothly over time. As the period shortens to $l = 4\pi$, more frequent slope changes intensify shear and enhance the diffusion-induced electric field, which couples the ionic species more strongly. This leads to a wider concentration band for species 1 and 3, while species 2 remains confined, and the effective diffusivity of species 1 exhibits non-monotonic temporal variation due to competing effects of shear, confinement, and electro-diffusive coupling. At the shortest period ($l = 2\pi$), rapid geometric variations generate steep velocity gradients and a pronounced induced potential, driving strong cross-species interactions and complex spreading patterns. Here, species 1 shows the broader range distribution, species 2 becomes less localized, and all effective diffusivities display dynamic, non-monotonic time evolution \figurename~\ref{15f}((g)-(i)), shows an initial enhancement followed by decay. As $l$ decreases, the transport regime shifts from diffusion dominated regime to the zone where dispersion is actively controlled by a combination of shear and electro-diffusive coupling. This interplay, mediated by channel geometry and the induced electric field, which enables advanced manipulation of ion transport and separation beyond classical Taylor-Aris dispersion.
	\subsubsection{Triangular wave channels with varying period length}
\begin{figure}
		\centering
		
		\begin{minipage}[t]{0.32\textwidth}
			\centering
			\begin{minipage}[t]{0.1\textwidth}
				\vspace{0pt}
				\textbf{(a)}
			\end{minipage}%
			\begin{minipage}[t]{0.9\textwidth}
				\vspace{0.5cm}
				\includegraphics[width=\linewidth]{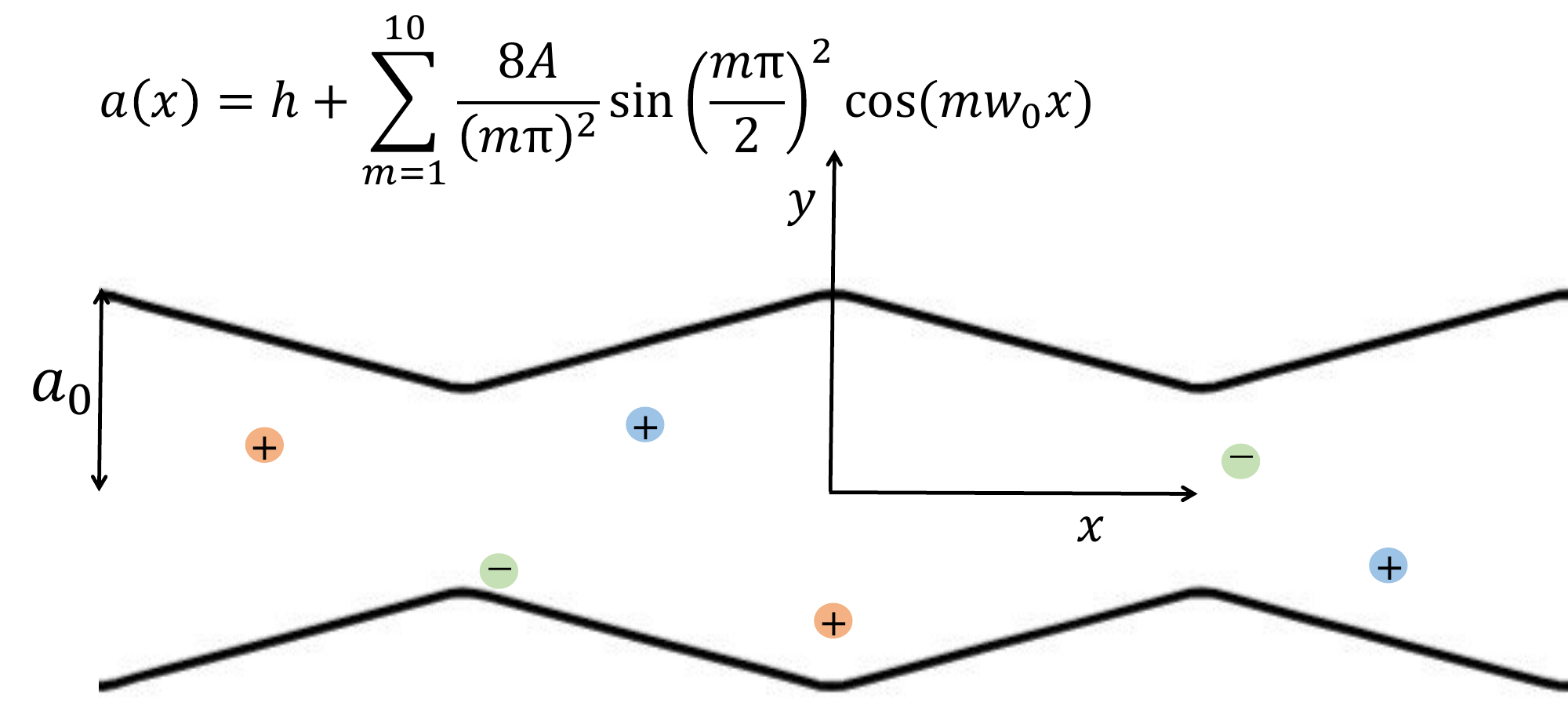}
			\end{minipage}
		\end{minipage}%
		\hfill
		\begin{minipage}[t]{0.32\textwidth}
			\centering
			\begin{minipage}[t]{0.1\textwidth}
				\vspace{0pt}
				\textbf{(b)}
			\end{minipage}%
			\begin{minipage}[t]{0.9\textwidth}
				\vspace{0.5cm}
				\includegraphics[width=\linewidth]{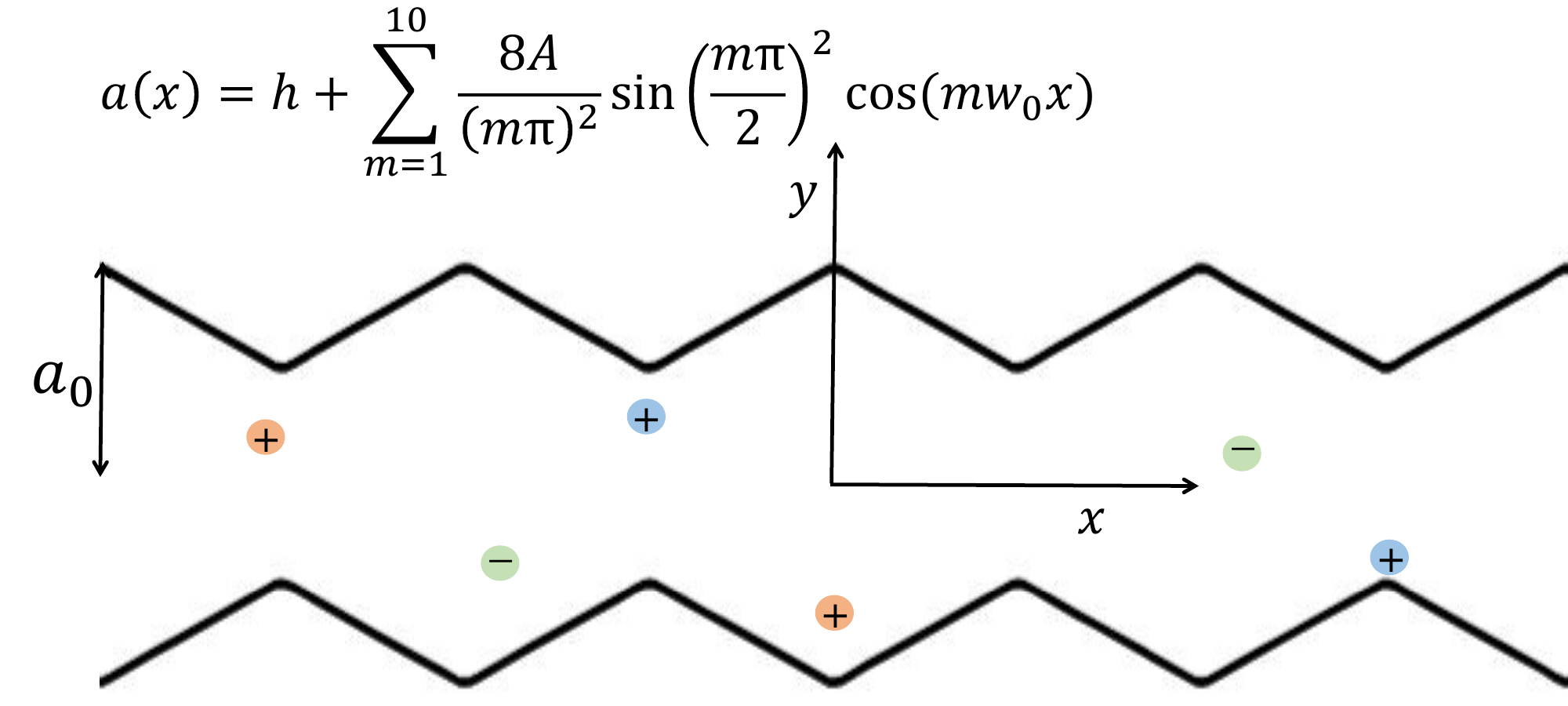}
			\end{minipage}
		\end{minipage}%
		\hfill
		\begin{minipage}[t]{0.32\textwidth}
			\centering
			\begin{minipage}[t]{0.1\textwidth}
				\vspace{0pt}
				\textbf{(c)}  
			\end{minipage}%
			\begin{minipage}[t]{0.9\textwidth}
				\vspace{0.5cm}
				\includegraphics[width=\linewidth]{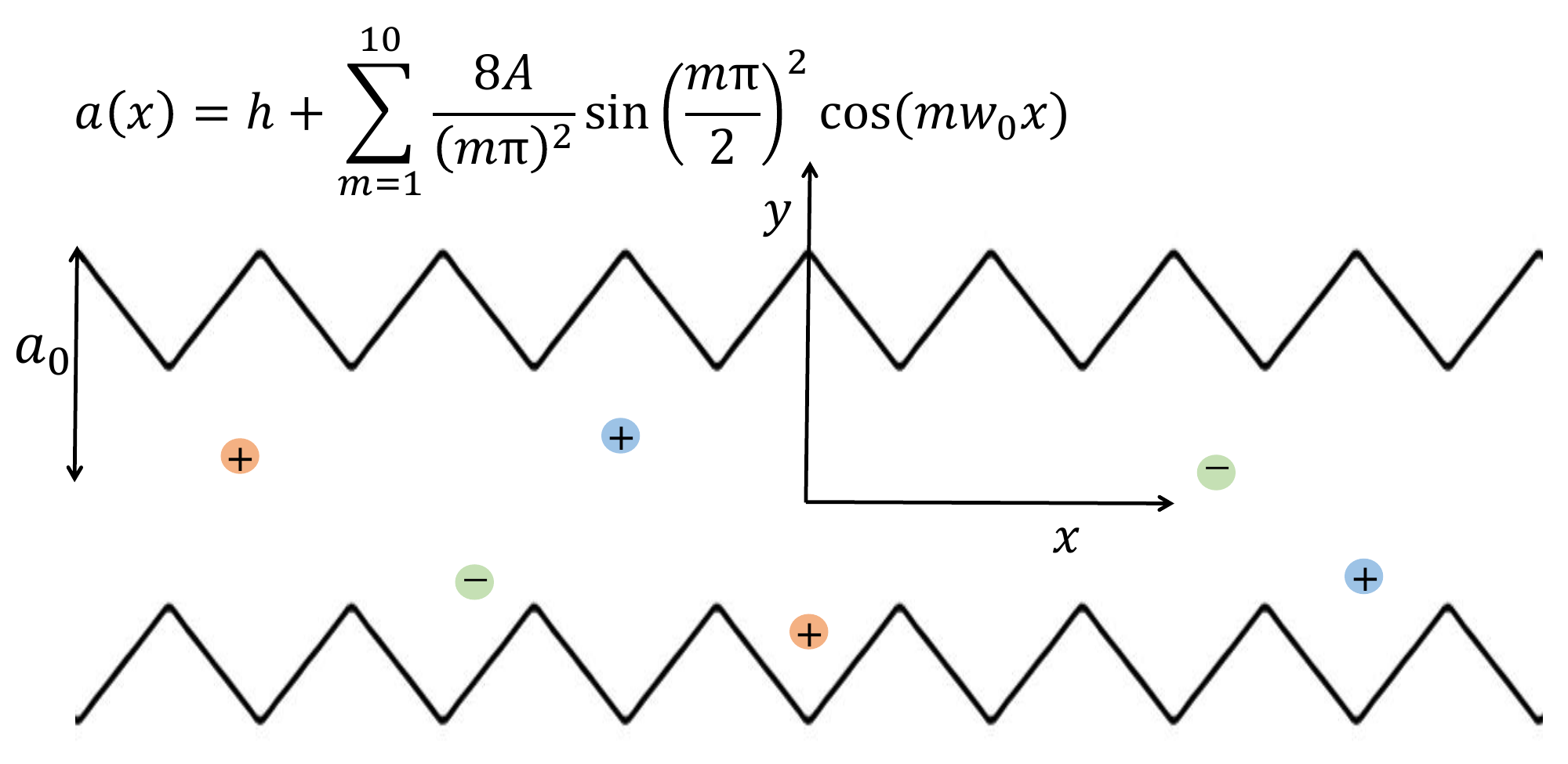}
			\end{minipage}
			\vspace{0.5cm}
		\end{minipage}
		\vspace{0.5cm}
		\begin{minipage}[t]{0.32\textwidth}
			\centering
			\begin{minipage}[t]{0.1\textwidth}
				\vspace{0pt}
				\textbf{(d)}
			\end{minipage}%
			\begin{minipage}[t]{0.9\textwidth}
				\vspace{0pt}
				\includegraphics[width=\linewidth]{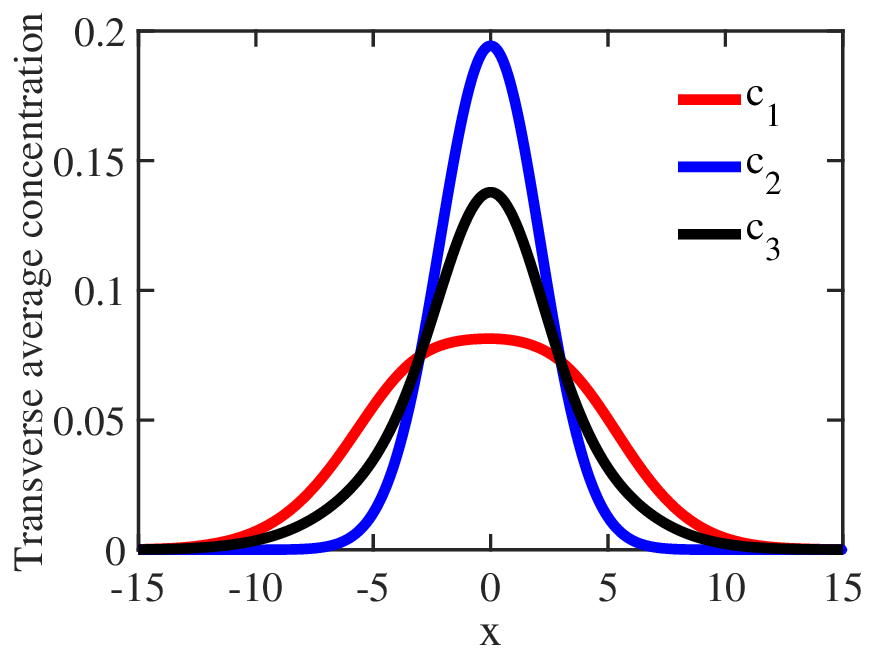}
			\end{minipage}
		\end{minipage}%
		\hfill
		\begin{minipage}[t]{0.32\textwidth}
			\centering
			\begin{minipage}[t]{0.1\textwidth}
				\vspace{0pt}
				\textbf{(e)}
			\end{minipage}%
			\begin{minipage}[t]{0.9\textwidth}
				\vspace{0pt}
				\includegraphics[width=\linewidth]{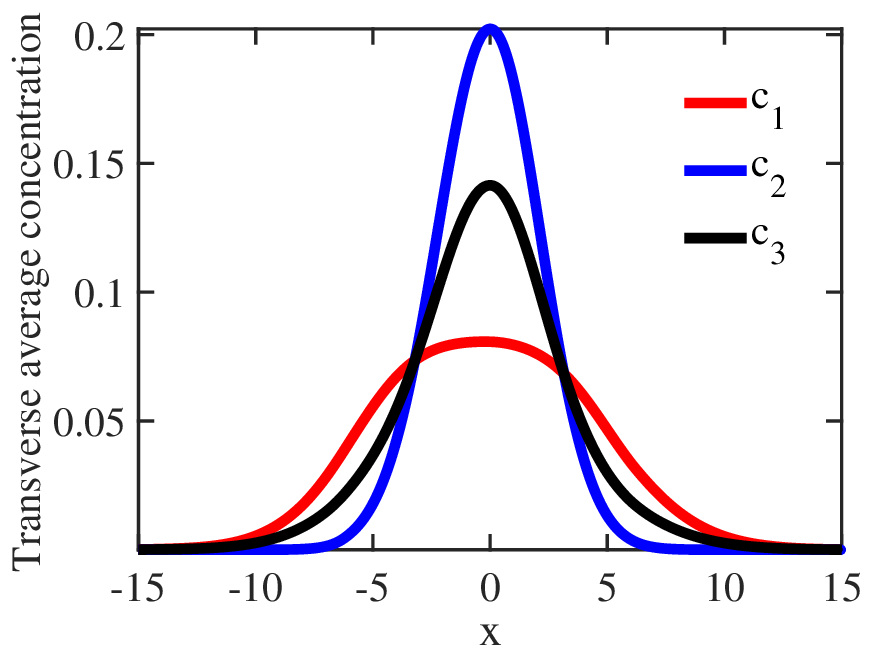}
			\end{minipage}
		\end{minipage}%
		\hfill
		\begin{minipage}[t]{0.32\textwidth}
			\centering
			\begin{minipage}[t]{0.1\textwidth}
				\vspace{0pt}
				\textbf{(f)}  
			\end{minipage}%
			\begin{minipage}[t]{0.9\textwidth}
				\vspace{0pt}
				\includegraphics[width=\linewidth]{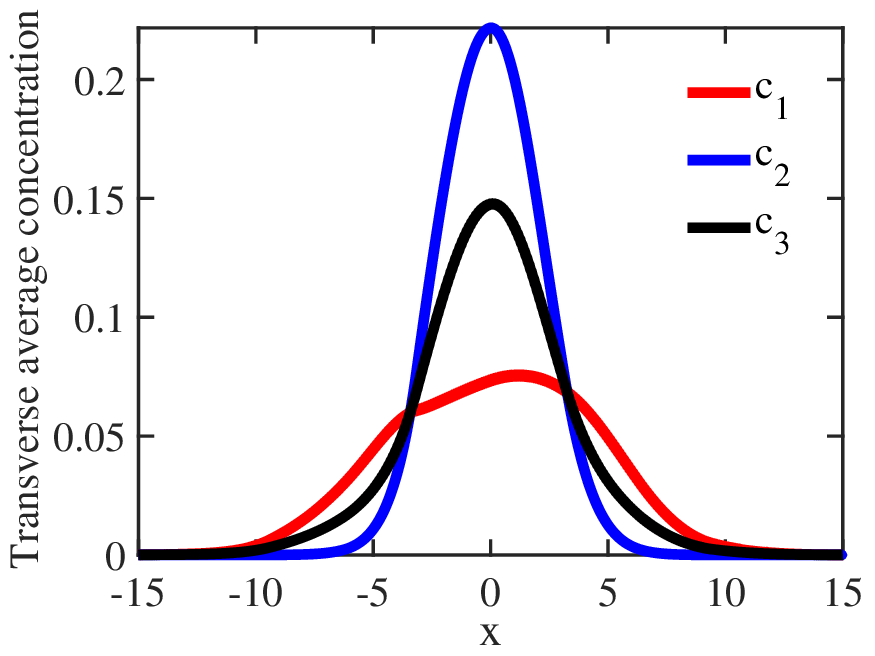}
			\end{minipage}
		\end{minipage}
		\hfill
		\begin{minipage}[t]{0.32\textwidth}
			\centering
			\begin{minipage}[t]{0.1\textwidth}
				\vspace{0pt}
				\textbf{(g)}
			\end{minipage}%
			\begin{minipage}[t]{0.9\textwidth}
				\vspace{0pt}
				\includegraphics[width=\linewidth]{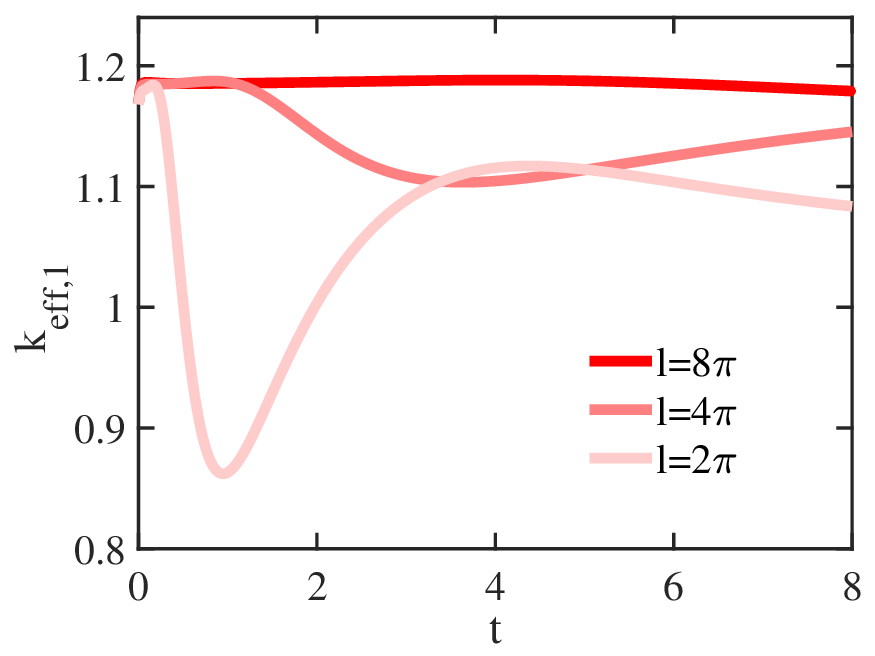}
			\end{minipage}
		\end{minipage}%
		\hfill
		\begin{minipage}[t]{0.32\textwidth}
			\centering
			\begin{minipage}[t]{0.1\textwidth}
				\vspace{0pt}
				\textbf{(h)}
			\end{minipage}
			\begin{minipage}[t]{0.9\textwidth}
				\vspace{0pt}
				\includegraphics[width=\linewidth]{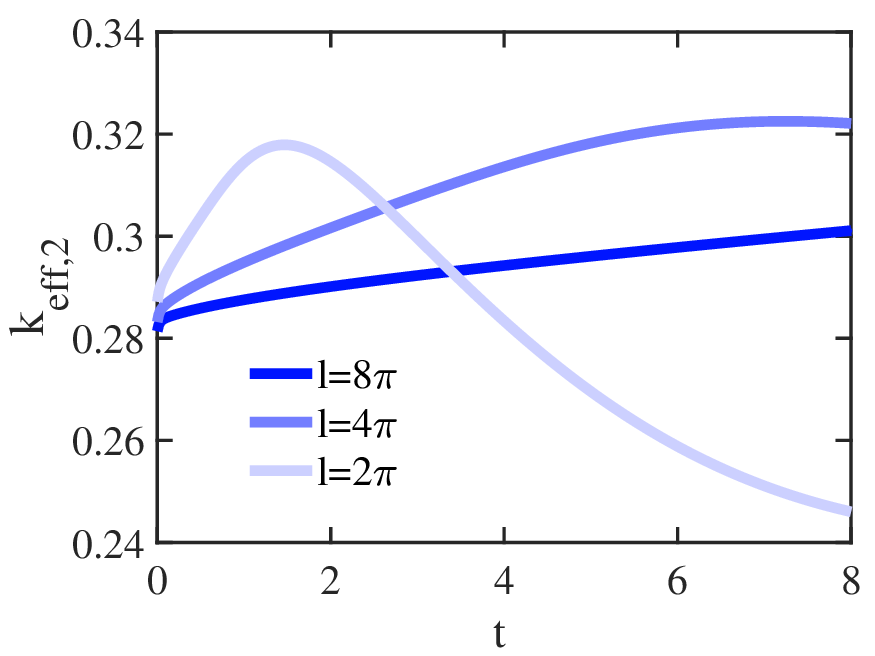}
			\end{minipage}
		\end{minipage}%
		\hfill
		\begin{minipage}[t]{0.32\textwidth}
			\centering
			\begin{minipage}[t]{0.1\textwidth}
				\vspace{0pt}
				\textbf{(i)}  
			\end{minipage}%
			\begin{minipage}[t]{0.9\textwidth}
				\vspace{0pt}
				\includegraphics[width=\linewidth]{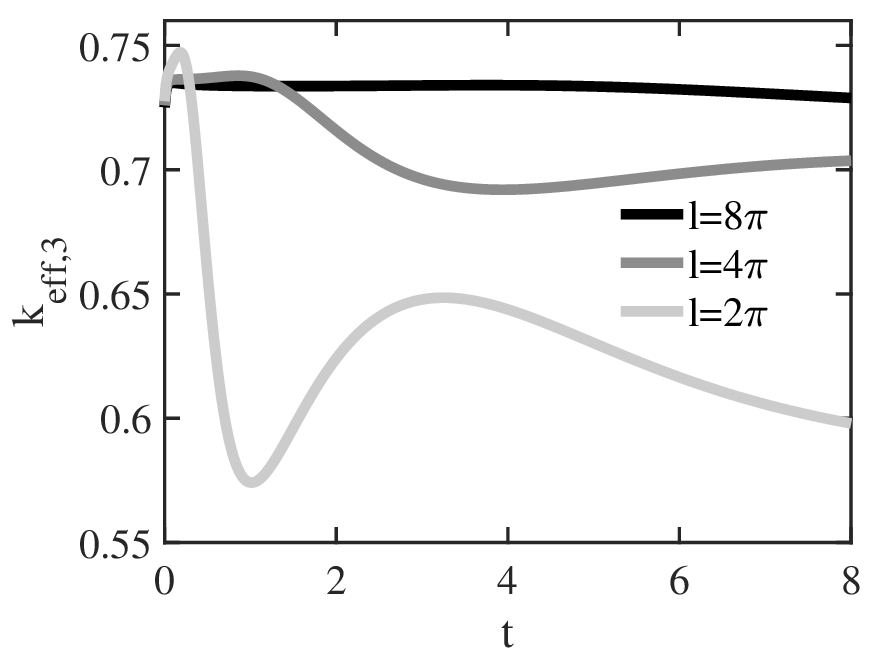}
			\end{minipage}
		\end{minipage}
		\caption{
			Schematic representation of the triangular wave channel geometry with different periodic lengths, $l$ (a) $l=8\pi$\;(two period), (b) $l=4\pi$\;(four period), and (c) $l=2\pi$\;(eight period), with the inlet height $a_{0}$  and amplitude $A=0.5$, where $\omega_{0}=\frac{2\pi}{l}$. (d-f) represents the Transverse-averaged concentration fields along the streamwise direction at time $t=8$ for $Pe=2$ and $Q=1$, corresponding to different periodic lengths. (g-i) represents the temporal evaluation of the effective diffusivities $k_{eff, i},\;i=1,2,3$ for three different ionic species corresponding to three different periodic lengths: $l=8\pi$, $l=4\pi$, and $l=2\pi$. 
		}  
		\label{18f}
	\end{figure}

	The mathematical formulation of the triangular wave patterning channels are framed by a Fourier series representation of a sawtooth profile, given by
	\begin{align*}  
		a(x) = h + \sum_{m=1}^{10} \frac{8A}{(m\pi)^2} \sin^2\left(\frac{m\pi}{2}\right) \cos(m \omega_0 x),
	\end{align*}
	where $h$ is the mean height, $A$ the amplitude, and $\omega_0 = 2\pi/l$ the fundamental frequency corresponding to the period length $l$. Compared to sinusoidal channels, the triangular geometry introduces sharp corners and abrupt changes in slope, leading to stronger localised shear and more pronounced velocity gradients. As the periodic length decreases from $l = 8\pi$ to $l = 2\pi$, the spatial frequency of these slope discontinuities increases, amplifying both shear-induced dispersion and the diffusion-induced electric field. The resulting transverse-averaged concentration profiles represented in \figurename~\ref{18f}(d-f), shows enhanced species separation and distinctly non-Gaussian spreading relative to sinusoidal cases at the same P\'eclet number and flow rate. In particular, the sharp transitions in triangular channels generate stronger confinement and stretching of concentration fields, which combined with electro-diffusive coupling yields species-dependent effective diffusivities that evolve non-monotonically in time which are represented in \figurename~\ref{18f}(g-i). It is observed that sinusoidal channels produce smooth, gradual variations in dispersion, but triangular geometries leverage abrupt geometric changes to achieve more efficient mixing and sharper focusing, making the separation efficient strategy suitable for applications where rapid separation or enhanced cross-stream transport is desired. This comparison underscores how geometric sharpness, not just periodic modulation, can be tailored to control multi-species dispersion in electrokinetic systems.    
	\subsubsection{For the converging channel geometries}
\begin{figure}
		\centering
		
		\begin{minipage}[t]{0.32\textwidth}
			\centering
			\begin{minipage}[t]{0.1\textwidth}
				\vspace{0pt}
				\textbf{(a)}
			\end{minipage}%
			\begin{minipage}[t]{0.9\textwidth}
				\vspace{0.5cm}
				\includegraphics[width=\linewidth]{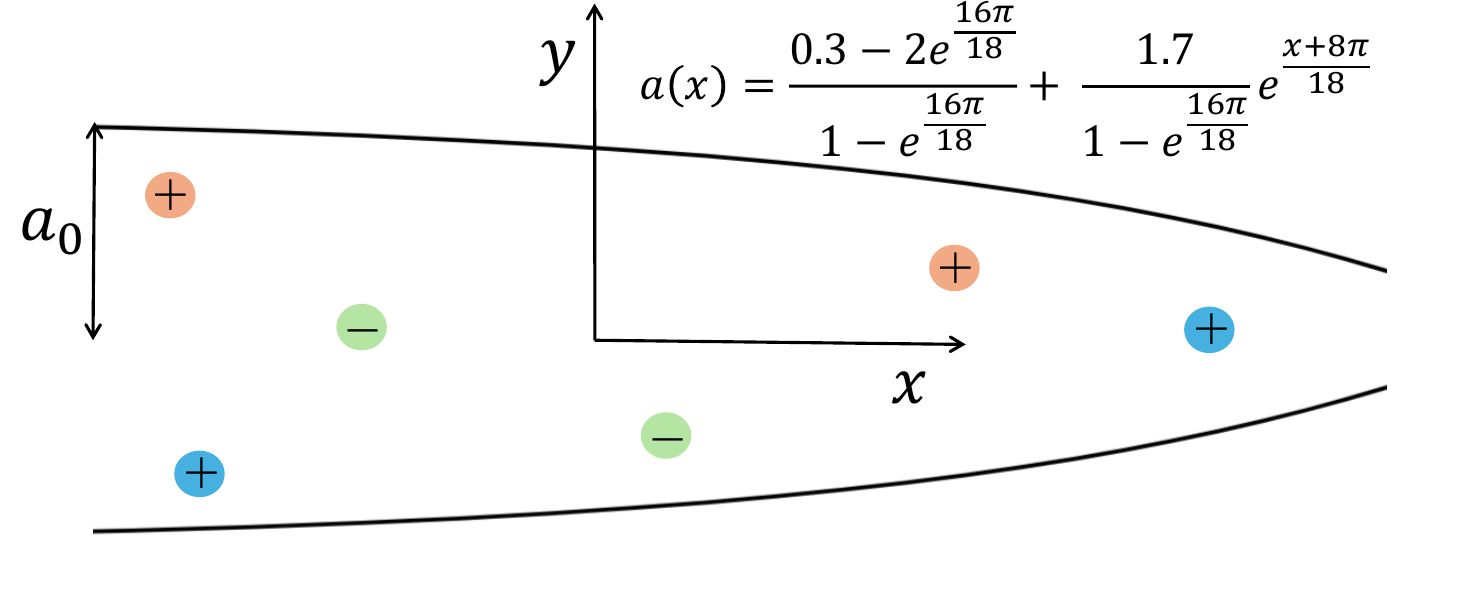}
			\end{minipage}
		\end{minipage}%
		\hfill
		\begin{minipage}[t]{0.32\textwidth}
			\centering
			\begin{minipage}[t]{0.1\textwidth}
				\vspace{0pt}
				\textbf{(b)}
			\end{minipage}%
			\begin{minipage}[t]{0.9\textwidth}
				\vspace{0.5cm}

				\includegraphics[width=\linewidth]{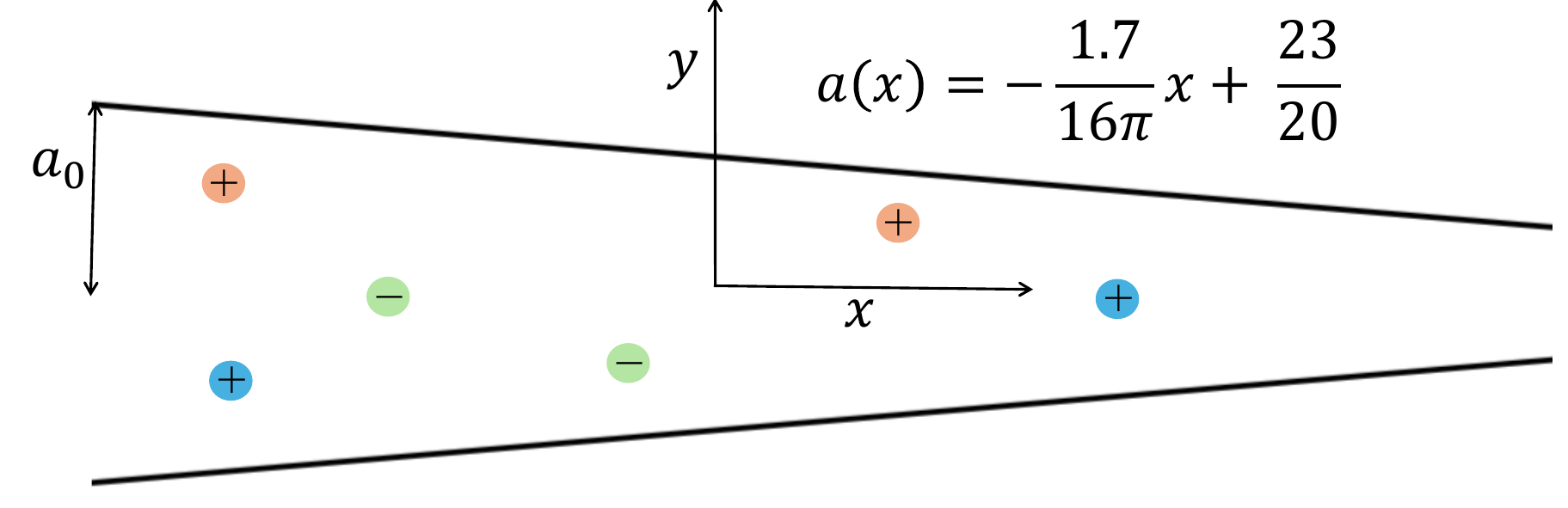}
				
			\end{minipage}
		\end{minipage}%
		\hfill
		\begin{minipage}[t]{0.32\textwidth}
			\centering
			\begin{minipage}[t]{0.1\textwidth}
				\vspace{0pt}
				\textbf{(c)}  
			\end{minipage}%
			\begin{minipage}[t]{0.9\textwidth}
				\vspace{0.5cm}
				\includegraphics[width=\linewidth]{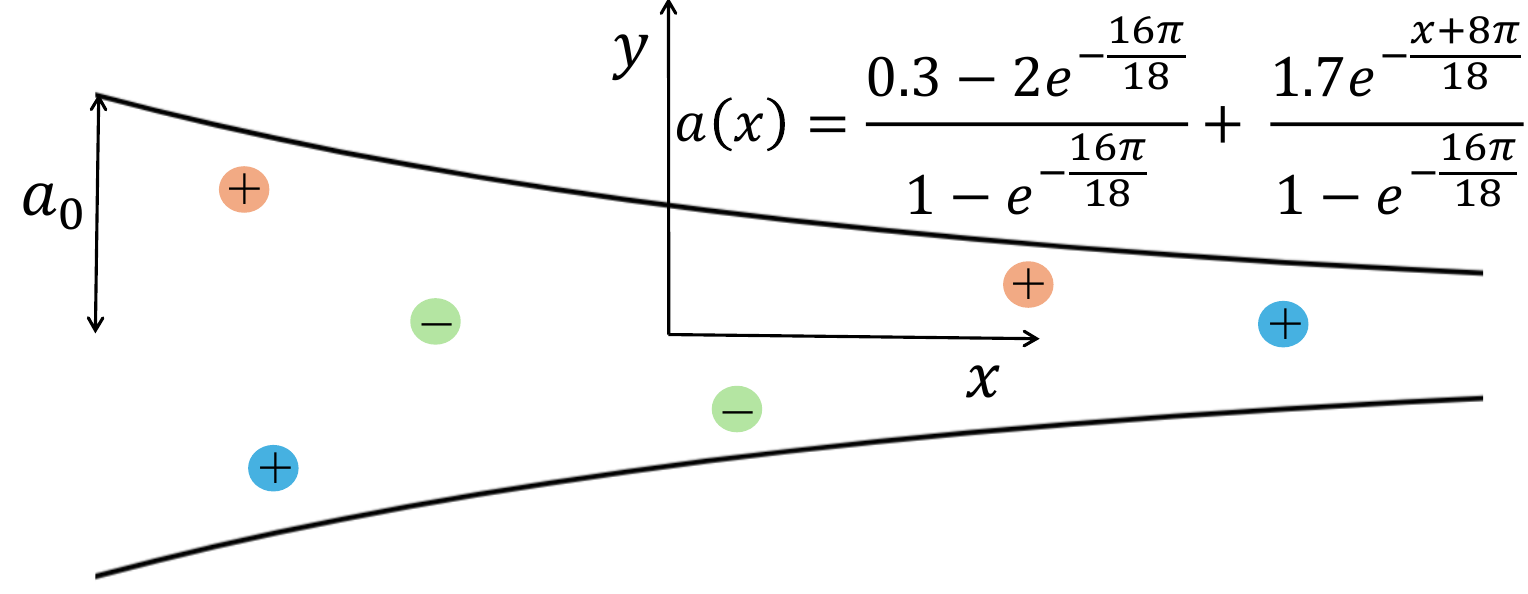}
			\end{minipage}
			\vspace{0.5cm}
		\end{minipage}
		\vspace{0.5cm}
		\begin{minipage}[t]{0.32\textwidth}
			\centering
			\begin{minipage}[t]{0.1\textwidth}
				\vspace{0pt}
				\textbf{(d)}
			\end{minipage}%
			\begin{minipage}[t]{0.9\textwidth}
				\vspace{0pt}
				\includegraphics[width=\linewidth]{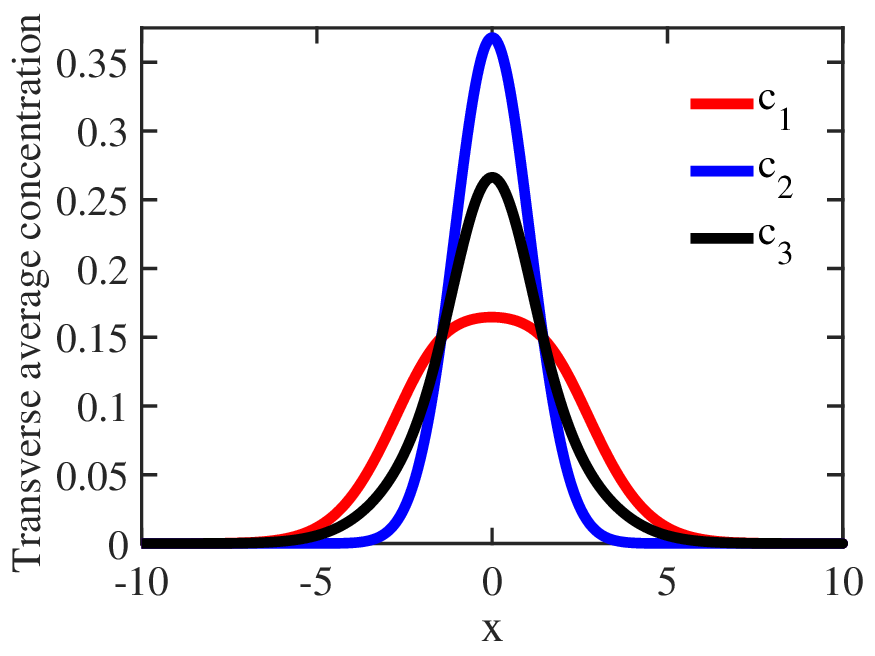}
			\end{minipage}
		\end{minipage}%
		\hfill
		\begin{minipage}[t]{0.32\textwidth}
			\centering
			\begin{minipage}[t]{0.1\textwidth}
				\vspace{0pt}
				\textbf{(e)}
			\end{minipage}%
			\begin{minipage}[t]{0.9\textwidth}
				\vspace{0pt}
				\includegraphics[width=\linewidth]{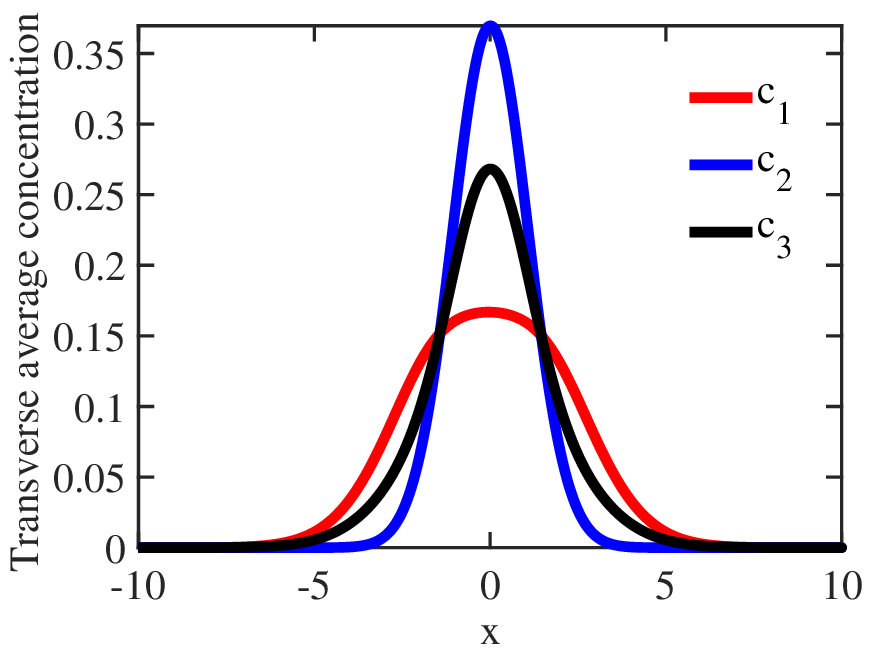}
			\end{minipage}
		\end{minipage}%
		\hfill
		\begin{minipage}[t]{0.32\textwidth}
			\centering
			\begin{minipage}[t]{0.1\textwidth}
				\vspace{0pt}
				\textbf{(f)}  
			\end{minipage}%
			\begin{minipage}[t]{0.9\textwidth}
				\vspace{0pt}
				\includegraphics[width=\linewidth]{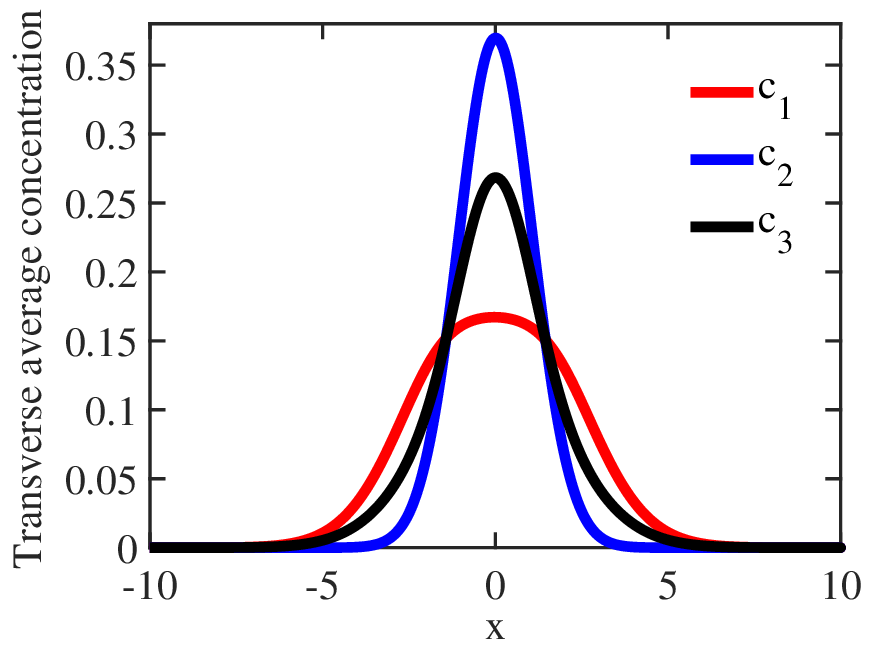}
			\end{minipage}
		\end{minipage}
		\hfill
		\begin{minipage}[t]{0.32\textwidth}
			\centering
			\begin{minipage}[t]{0.1\textwidth}
				\vspace{0pt}
				\textbf{(g)}
			\end{minipage}%
			\begin{minipage}[t]{0.9\textwidth}
				\vspace{0pt}
				\includegraphics[width=\linewidth]{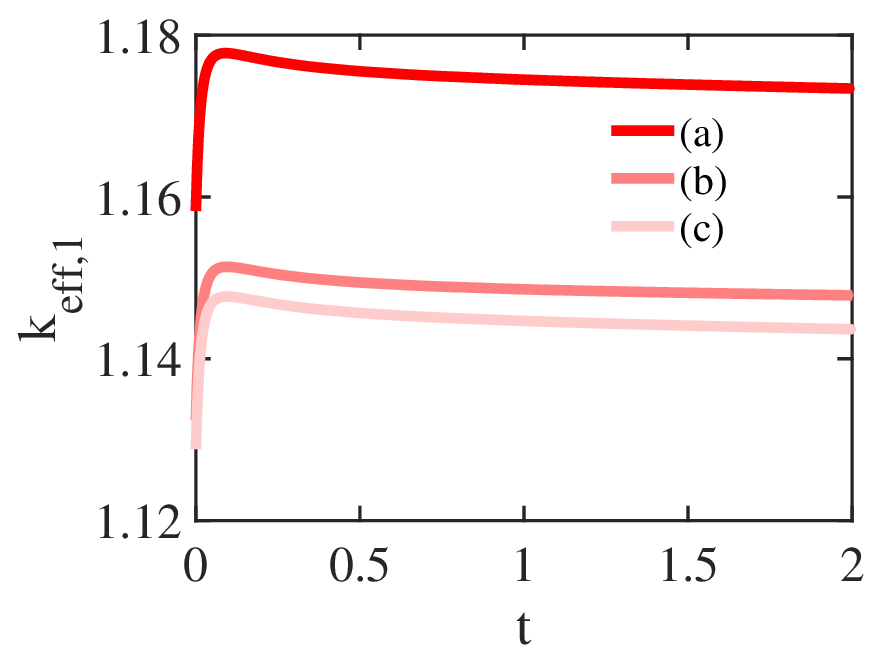}
			\end{minipage}
		\end{minipage}%
		\hfill
		\begin{minipage}[t]{0.32\textwidth}
			\centering
			\begin{minipage}[t]{0.1\textwidth}
				\vspace{0pt}
				\textbf{(h)}
			\end{minipage}%
			\begin{minipage}[t]{0.9\textwidth}
				\vspace{0pt}
				\includegraphics[width=\linewidth]{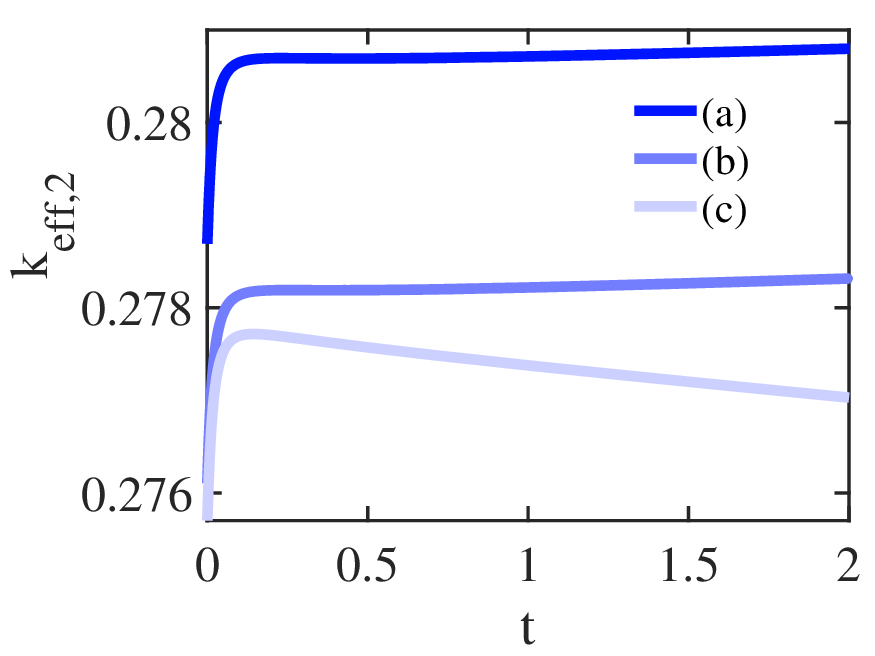}
			\end{minipage}
		\end{minipage}%
		\hfill
		\begin{minipage}[t]{0.32\textwidth}
			\centering
			\begin{minipage}[t]{0.1\textwidth}
				\vspace{0pt}
				\textbf{(i)}  
			\end{minipage}%
			\begin{minipage}[t]{0.9\textwidth}
				\vspace{0pt}
				\includegraphics[width=\linewidth]{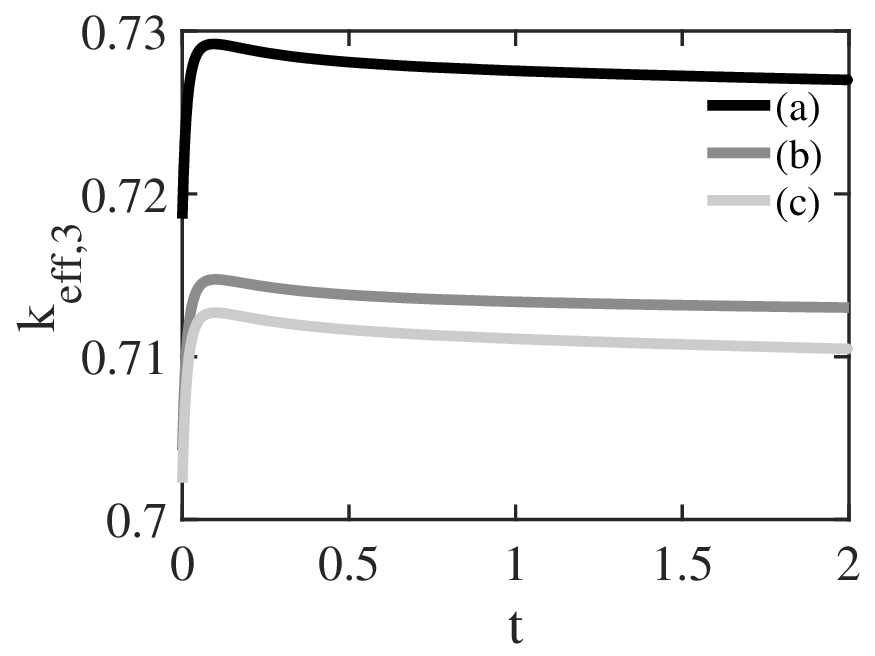}
			\end{minipage}
		\end{minipage}
		\caption{
			$(a)-(c)$ Represents the schematic of the concave downward converging channel, convergent nozzle, and concave upward converging channel with the inlet height $a_{0}$ and variable width $a(x)$. $(d)-(e)$ Represents the transverse averaged concentration distribution of three ionic species at time $t=2$ for $Pe=2$ and $Q=1$ within the different channel geometries, respectively. $(g)-(i)$ Represents the temporal evolution of the effective diffusivities $k_{eff, i}, i=1, 2, 3$ for three ion species in the concave downward diverging channel,  converging nozzle, and concave upward converging channel, respectively.
		}  
		\label{3f}
	\end{figure}

	In the concave downward converging channel \figurename~\ref{3f}(a), the narrowing geometry compresses the solute distribution, yielding sharper transverse averaged concentration profiles as compared to the straight channel. In \figurename~\ref{3f}(g)-(i), the effective diffusivities show pronounced asymmetry between the species where $k_{eff,1}$ and $k_{eff,3}$ decay monotonically with time, while $k_{eff,2}$ exhibits a gradual increase. This divergence indicates that the curvature modifies the hydrodynamic coupling between advection and diffusion in a species-dependent manner, thereby amplifying transport for one species while restricting it for others.
	In contrast, the concave upward converging channel \figurename~\ref{3f}(c) shows the concentration profiles that are broader than those in downward geometry but narrower than the straight channel. In \figurename~\ref{3f}(g)-(i), it can be easily predicted that the effective diffusivities of all three species decrease with time, with the most pronounced reduction observed for $k_{eff,2}$. This behavior signifies a more uniform suppression of dispersion across species, where the geometric expansion counteracts dispersive spreading and reduces the asymmetry as found in the downward case.
	A significantly different behavior emerges in the converging nozzle \figurename~\ref{3f}. The transverse averaged profiles are confined sharply near the centerline, with steep gradients across the channel cross-section. In \figurename~\ref{3f}(g)-(i), the effective diffusivities reveal contrasting dynamics: where $k_{eff,1}$ and $k_{eff,3}$ decreases steadily, consistent with confinement induced suppression, while $k_{eff,2}$ increases over time before approaching stabilization. This mixed response highlights the anisotropic influence of the nozzle geometry, which selectively enhances transport for one ionic species while inhibiting it for others.\\
	\figurename~\ref{3f} established a clear comparison across the converging geometries, in a straight channel, the effective dispersion coefficient $k_{eff}$ increases monotonically along the channel, consistent with the findings of \cite{ding2023shear}. However, this behaviour is significantly altered by geometric modulation. In convergent channels (concave downward, nozzle-shaped, and concave upward), the effective dispersion coefficients of the first and third species initially increase and then subsequently decreases, and eventually reach a constant value at long times, as shown in \figurename~\ref{3f}(g) and \figurename~\ref{3f}(h), even though the volumetric flow rate $Q$ remains constant. Under constant flow rate, the local fluid velocity increases in narrower cross-sections, where dispersion is found to be increased \cite{chang2023taylor, teodoro2025taylor}. However, in the present case, the electroneutrality condition forces all species to remain closely coupled, thereby suppressing relative spreading. As a result, electro-diffusive effects
	dominate over purely hydrodynamic dispersion, leading to a reduction in the effective dispersion coefficient and its eventual saturation at long times. For the second species, the effective dispersion coefficient remains nearly constant in the first two convergent geometries (\figurename~\ref{3f}(a) and \figurename~\ref{3f}(b)), owing to the gradual variation in cross-section and the relatively small molecular diffusivity of this species. In contrast, for the third convergent geometry (\figurename~\ref{3f}(c)), the dispersion coefficient decreases over time, following a similar trend to that observed for the first and third species. These results demonstrate that channel geometry is a key parameter for the solute transport, with curvature and nozzle effects introducing significant deviations from classical Taylor-Aris dispersion.
	\subsubsection{For the diverging channel geometries}
	\begin{figure}
		\centering
		\begin{minipage}[t]{0.32\textwidth}
			\centering
			\begin{minipage}[t]{0.1\textwidth}
				\vspace{0pt}
				\textbf{(a)}
			\end{minipage}%
			\begin{minipage}[t]{0.9\textwidth}
				\vspace{0.5cm}
				\includegraphics[width=\linewidth]{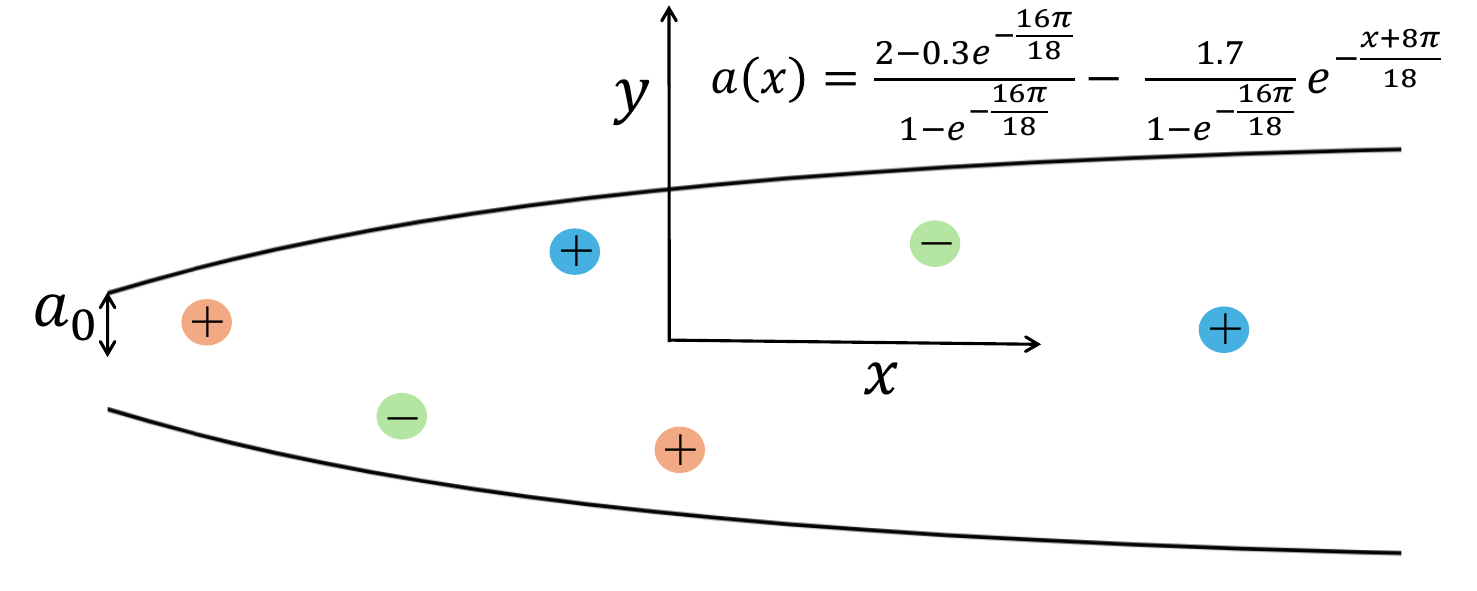}
			\end{minipage}
		\end{minipage}%
		\hfill
		\begin{minipage}[t]{0.32\textwidth}
			\centering
			\begin{minipage}[t]{0.1\textwidth}
				\vspace{0pt}
				\textbf{(b)}
			\end{minipage}%
			\begin{minipage}[t]{0.9\textwidth}
				\vspace{0.5cm}
				\includegraphics[width=\linewidth]{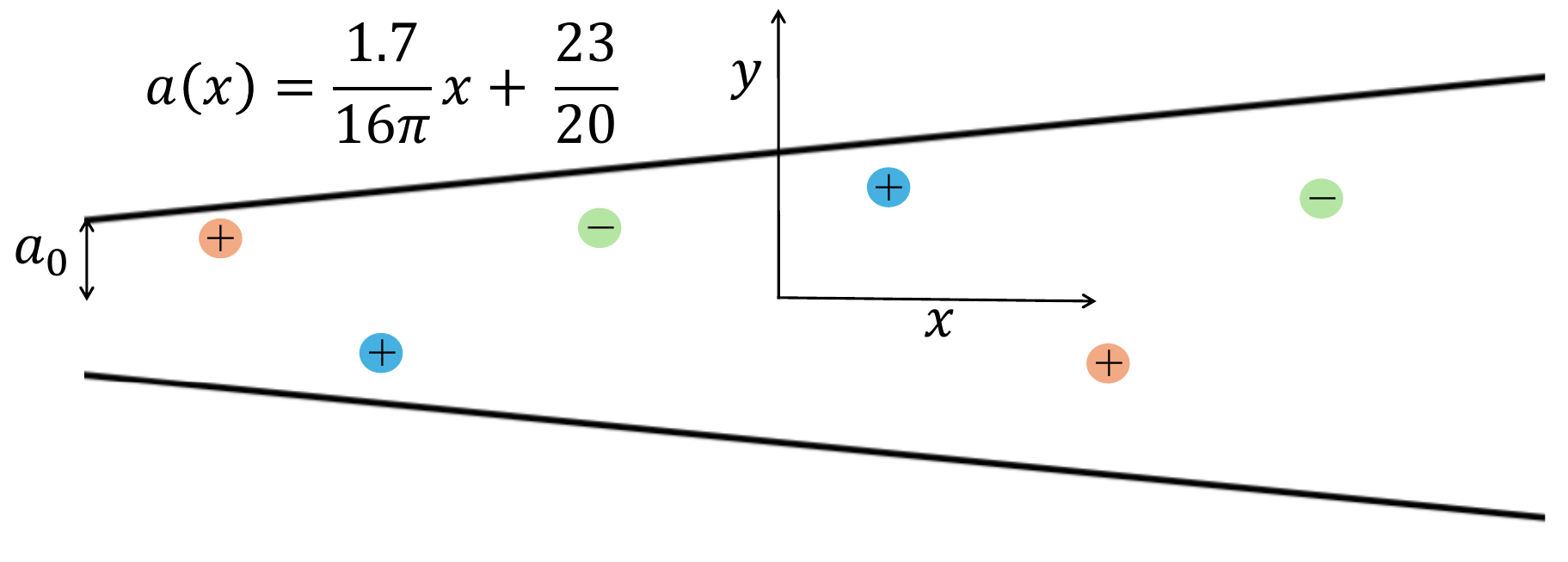}
				
			\end{minipage}
		\end{minipage}%
		\hfill
		\begin{minipage}[t]{0.32\textwidth}
			\centering
			\begin{minipage}[t]{0.1\textwidth}
				\vspace{0pt}
				\textbf{(c)}  
			\end{minipage}%
			\begin{minipage}[t]{0.9\textwidth}
				\vspace{0.5cm}

				\includegraphics[width=\linewidth]{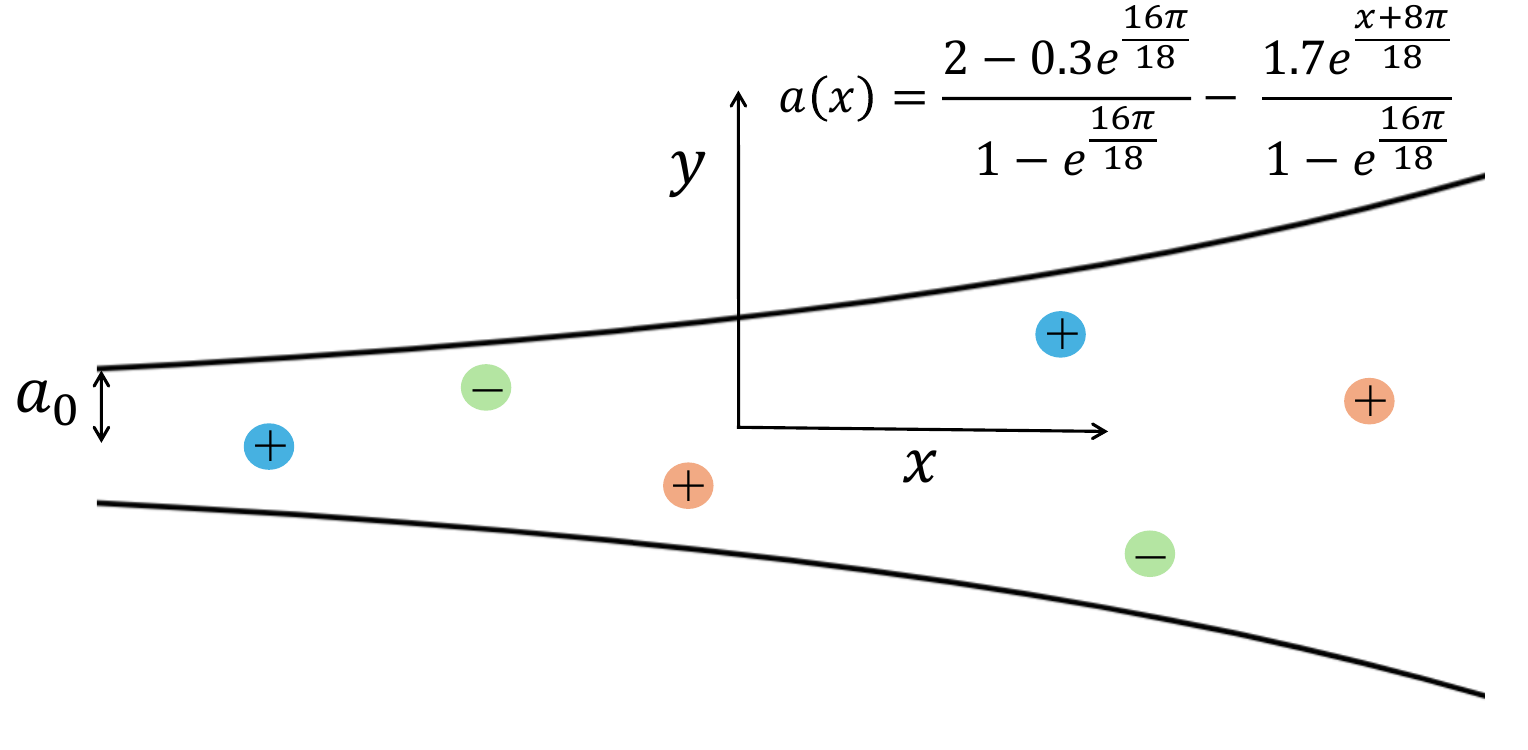}
			\end{minipage}
			\vspace{0.5cm}
		\end{minipage}
		\vspace{0.5cm}
		\begin{minipage}[t]{0.32\textwidth}
			\centering
			\begin{minipage}[t]{0.1\textwidth}
				\vspace{0pt}
				\textbf{(d)}
			\end{minipage}%
			\begin{minipage}[t]{0.9\textwidth}
				\vspace{0pt}
				\includegraphics[width=\linewidth]{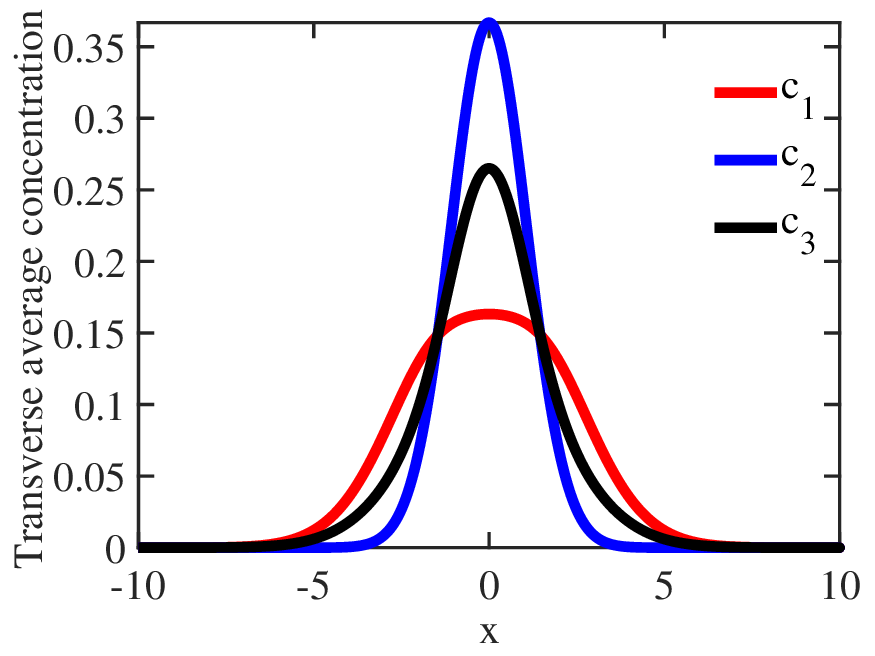}
			\end{minipage}
		\end{minipage}%
		\hfill
		\begin{minipage}[t]{0.32\textwidth}
			\centering
			\begin{minipage}[t]{0.1\textwidth}
				\vspace{0pt}
				\textbf{(e)}
			\end{minipage}%
			\begin{minipage}[t]{0.9\textwidth}
				\vspace{0pt}
				\includegraphics[width=\linewidth]{ 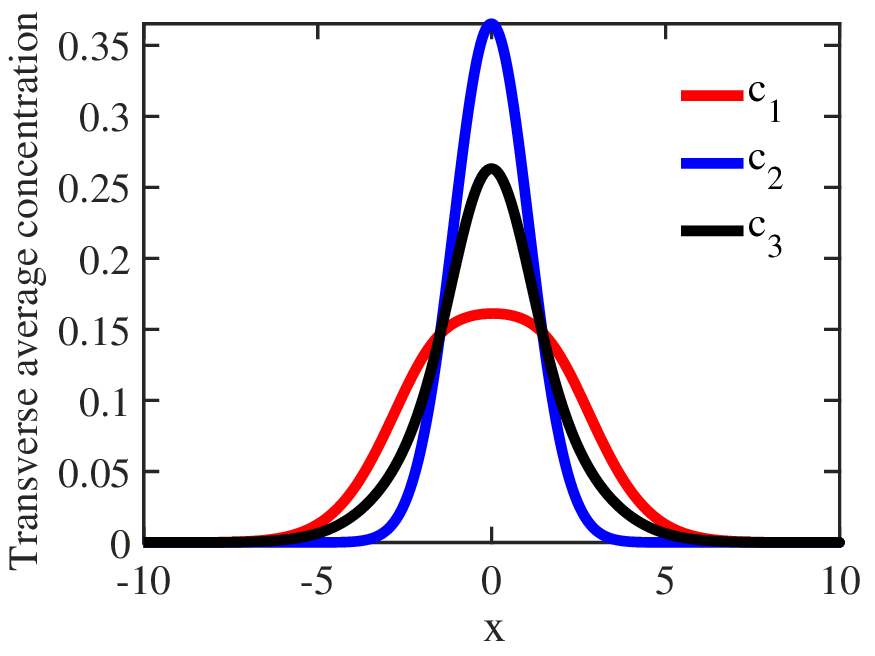}
			\end{minipage}
		\end{minipage}%
		\hfill
		\begin{minipage}[t]{0.32\textwidth}
			\centering
			\begin{minipage}[t]{0.1\textwidth}
				\vspace{0pt}
				\textbf{(f)}  
			\end{minipage}%
			\begin{minipage}[t]{0.9\textwidth}
				\vspace{0pt}
				\includegraphics[width=\linewidth]{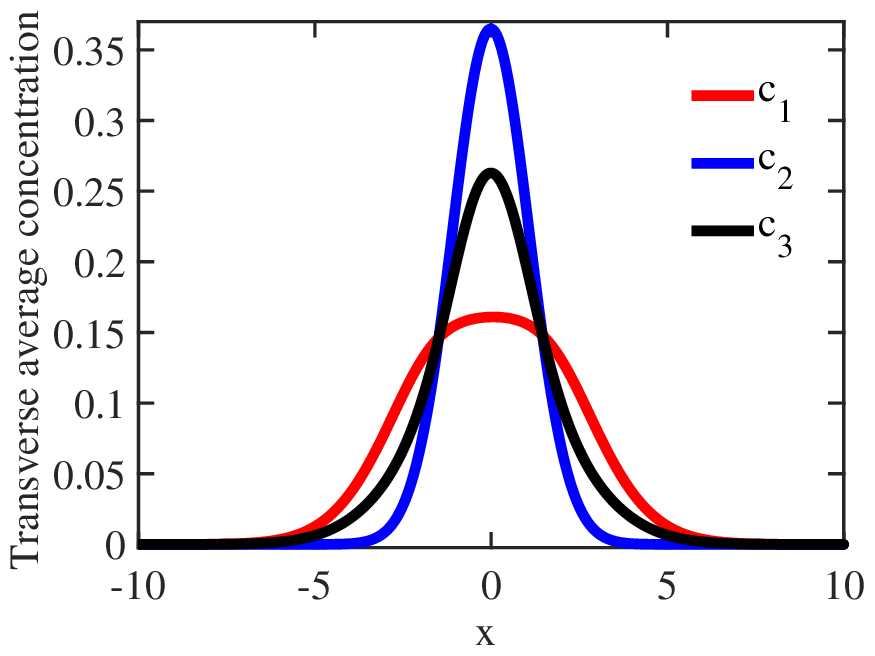}
			\end{minipage}
		\end{minipage}
		\hfill
		\begin{minipage}[t]{0.32\textwidth}
			\centering
			\begin{minipage}[t]{0.1\textwidth}
				\vspace{0pt}
				\textbf{(g)}
			\end{minipage}%
			\begin{minipage}[t]{0.9\textwidth}
				\vspace{0pt}
				\includegraphics[width=\linewidth]{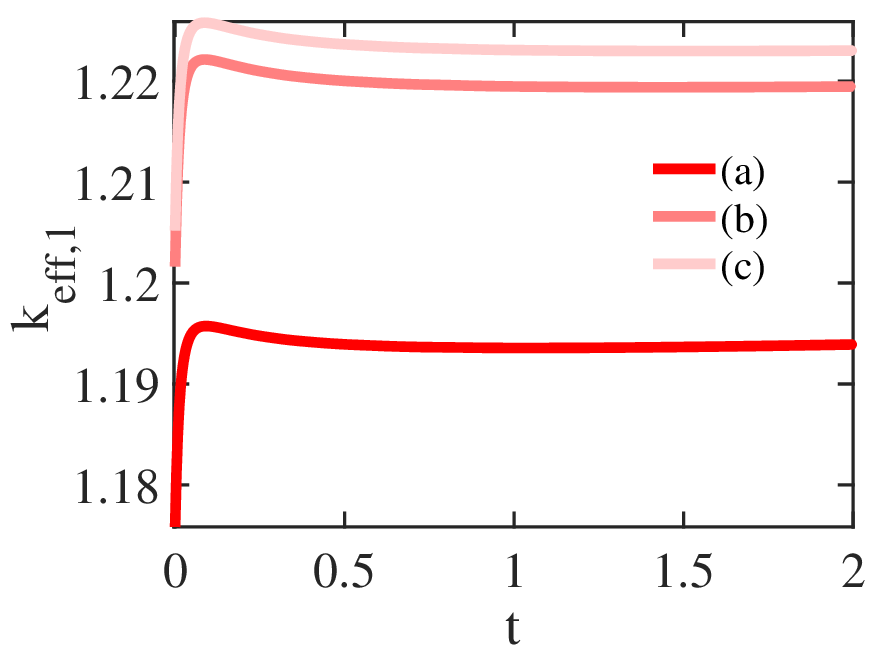}
			\end{minipage}
		\end{minipage}%
		\hfill
		\begin{minipage}[t]{0.32\textwidth}
			\centering
			\begin{minipage}[t]{0.1\textwidth}
				\vspace{0pt}
				\textbf{(h)}
			\end{minipage}%
			\begin{minipage}[t]{0.9\textwidth}
				\vspace{0pt}
				\includegraphics[width=\linewidth]{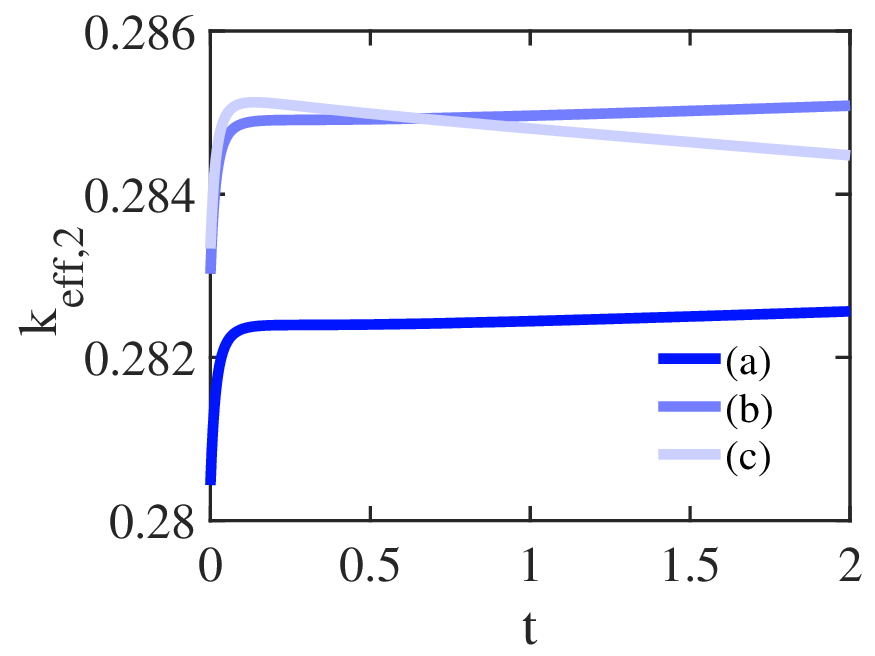}
			\end{minipage}
		\end{minipage}%
		\hfill
		\begin{minipage}[t]{0.32\textwidth}
			\centering
			\begin{minipage}[t]{0.1\textwidth}
				\vspace{0pt}
				\textbf{(i)}  
			\end{minipage}%
			\begin{minipage}[t]{0.9\textwidth}
				\vspace{0pt}
				\includegraphics[width=\linewidth]{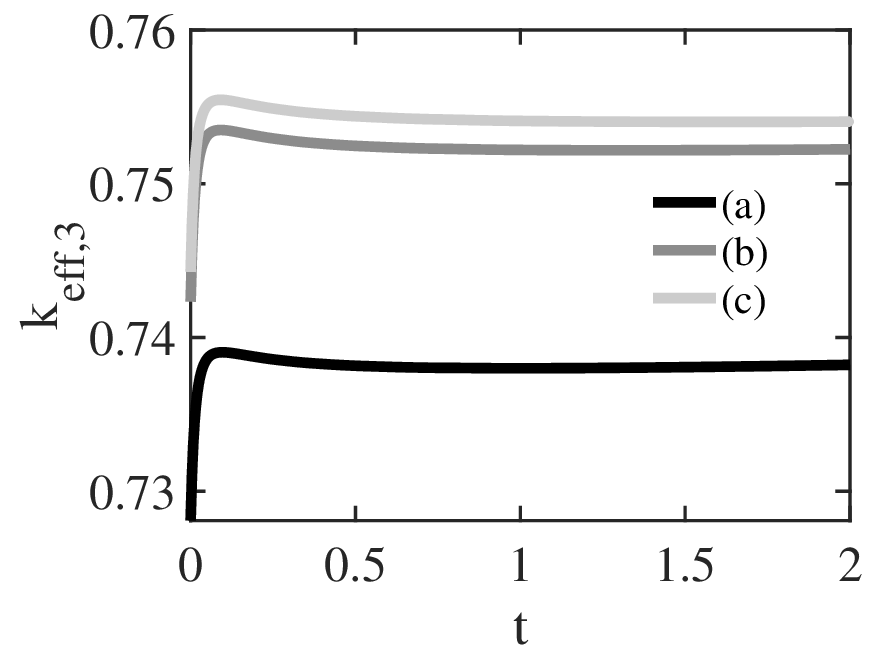}
			\end{minipage}
		\end{minipage}
		\caption{
			$(a)-(c)$ represents the schematic diagram of the concave downward diverging channel, diverging nozzle, and concave upward diverging channel, respectively. The channel height increases with the axial coordinate $x$, $a_{0}$ denotes the inlet height. $(d)-(f)$ Represents the transverse average of the concentration profiles for the three ion species $c_{i}$, $i = 1,2,3$, at time $t=2$, with $Pe=2$ and flow rate $Q=1$ within the different channel geometries, respectively. $(g)-(i)$ Represents the temporal evolution of the effective diffusivities $k_{eff, i}, i=1, 2, 3$ for three ionic species in the concave downward diverging channel, diverging nozzle, and concave upward converging nozzle, respectively. 
		}  
		\label{4f}
	\end{figure}
	In this section, the flow is assumed to be combined in a channel geometry where the wall patterning follows a concave downward pattern.
	In the concave downward diverging channel, the broadening cross-section allows enhanced lateral spreading of the solute represented in \figurename~\ref{4f}. Consequently, the transverse average concentration profiles are wider compared to converging geometries. Corresponding to the effective diffusivities ($k_{eff,1}$, $k_{eff,2}$, and $k_{eff,3}$) as shown in \figurename~\ref{4f} of all species exhibits an enhancement with respect to time, signifying that the expanding geometry amplifies longitudinal dispersion consistently across species. The effect is strongest for $k_{eff,2}$, while $k_{eff,1}$ and  $k_{eff,3}$ finds moderate growth. This confirms that the channel expansion augments dispersion and introduces species-dependent differences in transport rates.\\
	For the case of concave upward diverging patterns in \figurename~\ref{4f}, the solute distributions are again wider than in the straight channel, but broadening is less pronounced than in the concave downward geometry. The corresponding effective diffusivities in \figurename~\ref{4f} for all the species decrease gradually with time, suggesting that the upward curvature counteracts the dispersion enhancement effect of divergence and instead yields a net suppression of transport.\\
	The diverging nozzle induces species-dependent dispersion due to the interplay between its non-uniform flow field and electro-diffusive coupling and presents an intermediate behavior between the two concave divergent channels. From \figurename~\ref{4f}, it can be clearly visualized that the effective diffusivities $k_{eff,1}$ and $k_{eff,2}$ rise slightly over time as shear-enhanced dispersion acts, and the effect is moderated by nonlinear induced electric forces. In contrast, $k_{eff,3}$ remains constant, as its transport is more constrained by the electromigration required to maintain zero current, making it less sensitive to flow-induced spreading. This highlights the influence of geometric expansion on species distribution within the channel. Instead of uniformly enhancing dispersion, the expansion induces anisotropic redistribution of the species. Consequently, the dispersion process depends on the individual diffusivities and charges of the species, leading to non-uniform spreading across the channel.
	\subsubsection{Separation efficiency}
\begin{figure}
		\centering
		\includegraphics[width=0.6\linewidth]{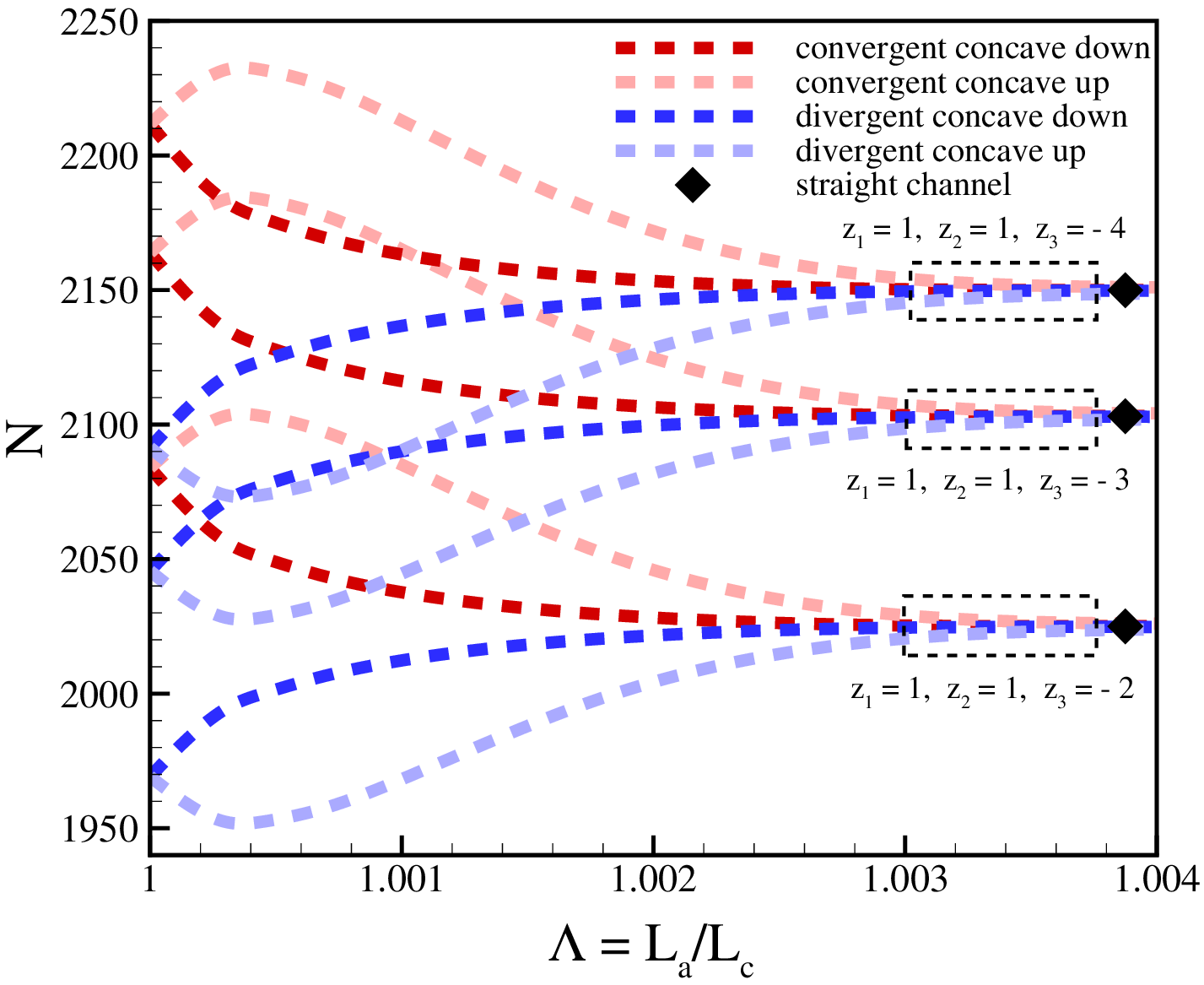}
		\caption{ Shows the number of theoretical plates, $N = \frac{L^2}{Var(c_{i})}$, as a function of the ratio of arc length to chord length $\Lambda$ for the species  $c_{1}$ with the different sets of valency $z = \{1, 1, -2\}$, $z = \{1, 1, -3\}$, $z = \{1, 1, -4\}$ in concave-up and concave-down convergent and divergent channel geometries.
		}
		\label{12fig1}
	\end{figure}

\begin{figure}
		\centering
	
		\includegraphics[width=0.6\linewidth]{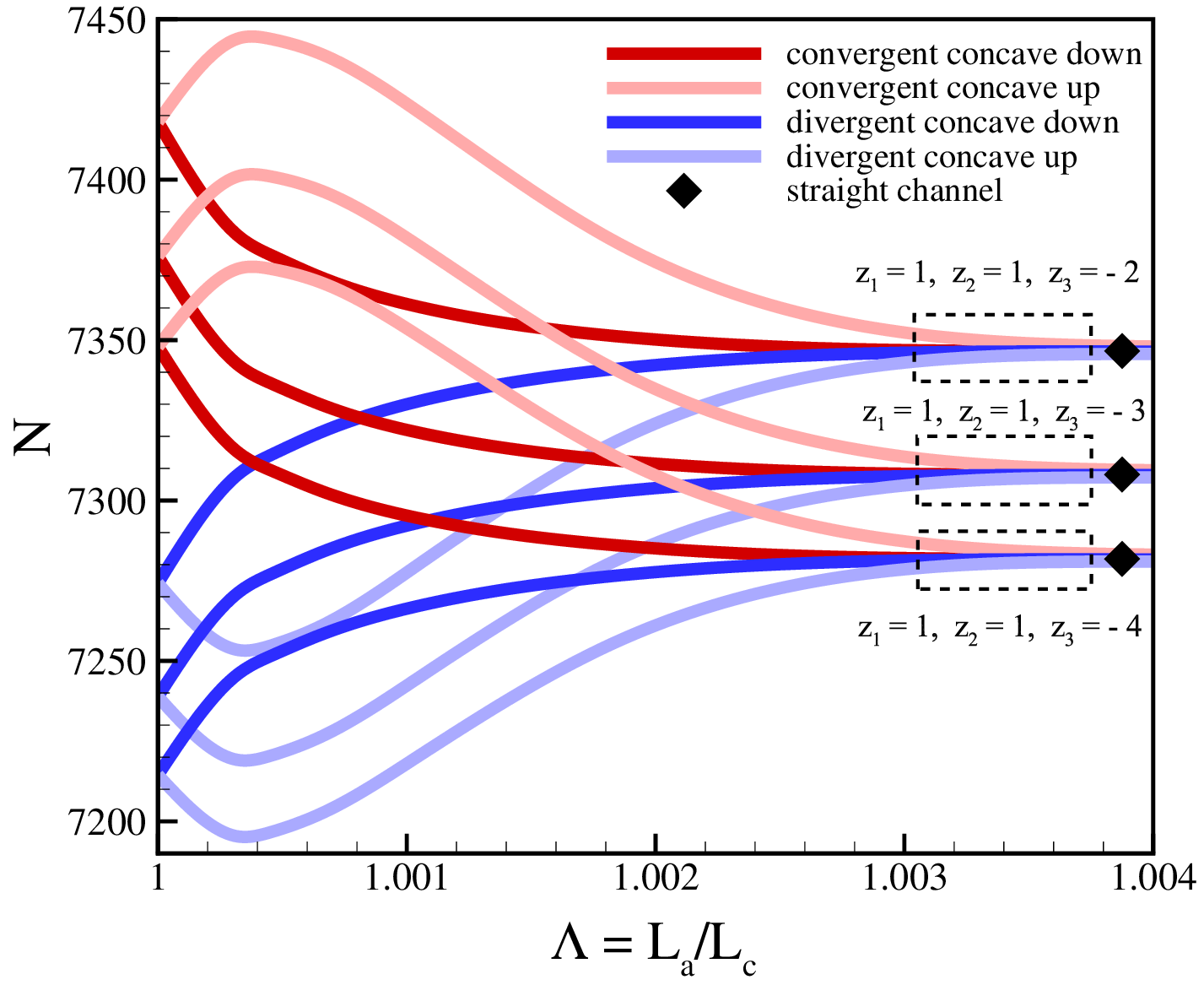}
		
		\caption{Shows the number of theoretical plates, $N = \frac{L^2}{Var(c_{i})}$, as a function of the ratio of arc length to chord length $\Lambda$ for the species $c_{2}$ with the same set of valances as shown in \figurename~\ref{12fig1}.
		}
		\label{12fig2}
	\end{figure}

\begin{figure}
		\centering
		
		\begin{minipage}[t]{0.5\textwidth}
			\centering
			\begin{minipage}[t]{0.1\textwidth}
				\vspace{0pt}
				\textbf{(a)}
			\end{minipage}%
			\begin{minipage}[t]{0.9\textwidth}
				\vspace{0pt}
				\includegraphics[width=\linewidth]
				{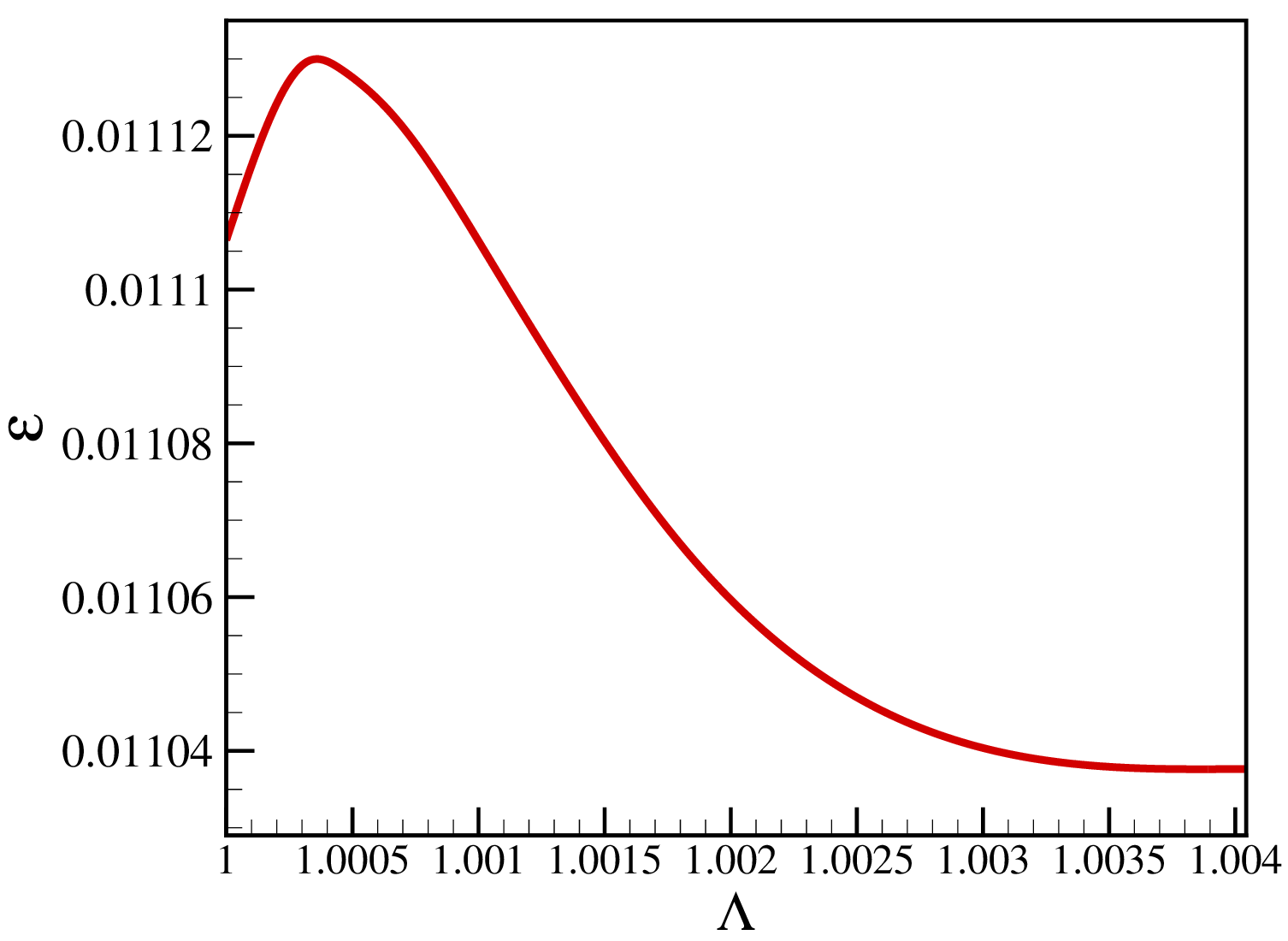}
			\end{minipage}
		\end{minipage}%
		\hfill
		\begin{minipage}[t]{0.5\textwidth}
			\centering
			\begin{minipage}[t]{0.1\textwidth}
				\vspace{0pt}
				\textbf{(b)}
			\end{minipage}%
			\begin{minipage}[t]{0.9\textwidth}
				\vspace{0pt}
				\includegraphics[width=\linewidth]{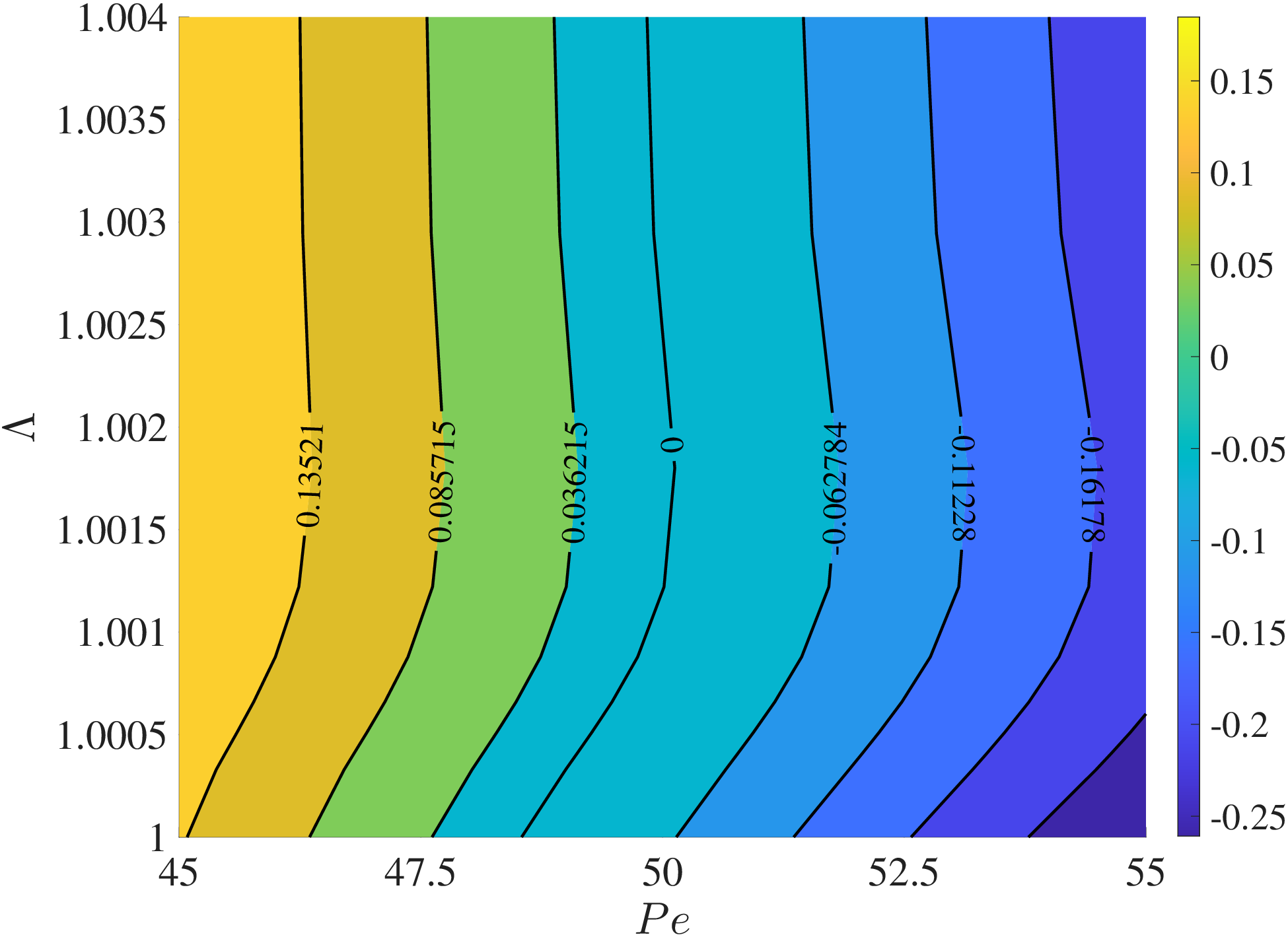}
			
			\end{minipage}
		\end{minipage}%
		\caption{ (a) shows the cumulative effect as a function of the $\Lambda$ and (b) Contour plot of the difference in effective diffusivities, $k_{eff,1}-k_{eff,2}$, as a function of the Péclet number $Pe$ and with the ratio parameter $\Lambda$. The color bar indicates the magnitude of $k_{eff,1}-k_{eff,2}$, with positive and negative regions and identifying the regimes where each species exhibits dominant behaviour.
		}
		\label{Figure 9}
	\end{figure}

	In capillary electrophoresis, the separation efficiency is typically measured by the 'number of theoretical plates' $N=\frac{L^2}{Var(c_i)}$, a dimensionless parameter which characterizes peak broadening\cite{ghosal2012electromigration}, where  
	\begin{align}
		Var(c_i) = \int_{-\infty}^{\infty} c_{i}x^2dx - \left(\int_{-\infty}^{\infty} c_{i}xdx\right)^2
	\end{align}
	is the variance of the transverse averaged concentration field $c_{i}$.
	To quantify the channel curvature, we introduce the geometric parameter $\Lambda = \frac{L_a}{L_c}$, defined as the ratio of the arc length to the chord length, where  $L_a$ is given by
	\begin{align}
		L_a = \int_{-l}^{l}\sqrt{1+\left(\frac{da}{dx}\right)^2} dx.
	\end{align}
	\figurename~\ref{12fig1} and \figurename~\ref{12fig2} illustrates the dependence of $N$ on this parameter. The number of theoretical plates corresponding to the first and second ion species in the convergent concave upward channel exhibits a non-monotonic behaviour. $N$ attains its maximum value for an intermediate value of $\Lambda$, indicating the optimal geometric configuration for separation. The peak value of $N$ further depends on the chosen ionic set of valencies, highlighting the coupled influence of the slope of the wall geometry and the diffusion-induced electric potential. The dependency on $\Lambda$ is non-monotonic with $N$ and within the convergent concave upward geometry, $N$ increases at the beginning with $\Lambda$, but then started decline after some intermediate value of $\Lambda$. \\
	To understand the origin of the non-monotonic behaviour of the separation efficiency, a parameter referred as the cumulative effect, denoted by $\mathcal{E}$ is introduced. This parameter measures the time-integrated influence of the induced electric potential on solute transport, rather than its instantaneous contribution to the dispersion. Physically, $\mathcal{E}$ accounts for the electric potential generated by differences in species diffusivities in the presence of spatial concentration gradients and is defined by
	\begin{align}
		\mathcal{E} = \int_{0}^{t}\left(\frac{1}{2l}\int_{-l}^{l}\left|\frac{\partial\phi}{\partial x}\right| dx\right)dt.
	\end{align}
	\figurename~\ref{Figure 9} shows the variation of the cumulative effect $\mathcal{E}$ with $\Lambda$ for the first ionic species and reveals a clear non-monotonic dependence. Notably, $\mathcal{E}$ attains a maximum value at a particular value of $\Lambda$, which coincides with the value at which the separation efficiency is maximized. At this optimal ratio, the enhanced cumulative electric effect suppresses the longitudinal spreading, leading to a minimum effective diffusivity of the first species, as shown in Figure S4(a) of the supplementary material.
	Further insight is obtained from Figure S4(b) of the supplementary material, which shows that the peak of the transversely averaged concentration of the first species initially increases as $\Lambda$ approaches to a critical value, indicating a reduction in variance but subsequently decreases with increase of the value of $\Lambda$, corresponding to an increase in variance associated with enhanced longitudinal spreading. The interplay between these competing mechanisms leads to a maximum separation efficiency at an intermediate value of $\Lambda$, close to one.   
	\subsubsection{ Dependency of the effective diffusivity on Peclet number}
	In this section, examine the dependence of the effective diffusivity on the Peclet number for a convergent concave-upward channel, corresponding to the value $\Lambda$ at which $N$ attains its maximum. As illustrated in Figure S3(b), the effective diffusivity varies significantly with the Peclet number. It is observed that, in contrast to classical Taylor dispersion theory, the presence of diffusion-induced electric potential leads to a non-monotonic variation of the effective diffusivity of the first ion species (red solid line) with respect to the Peclet number. Moreover, the ion species exhibiting the largest effective diffusivity depends on the Peclet number. At low Peclet numbers, the first ion species has the largest effective diffusivity. However, at $Pe\approx 48$, due to the combined effect of advection and the slope of the wall geometry, all three ion species exhibit the same effective diffusivity, in contrast to the results reported by \cite{ding2023shear}. For Large Peclet numbers, the second ion species (solid black line) exhibits the highest effective diffusivity. \figurename~\ref{Figure 9}(b) further shows that $k_{eff,1}-k_{eff,2}$ becomes zero, indicating that $k_{eff,1}$ and $k_{eff,2}$ intersect at different Peclet numbers for different geometries corresponding to a given value of $\Lambda$. The line which delineates the boundary of the effective diffusivity where $k_{eff,}-k_{eff,2}=0$, separating the regions of different species distribution predicting the dominant behaviour. In the regime where the difference is positive, $k_{eff,1}$ exhibits the highest effective diffusivity, and $k_{eff,2}$ represents the relative weak diffusivity. These observations demonstrate that the effective dispersion coefficient is controlled by the combined effects of the diffusion-induced electric potential and the channel geometry.
	
	\section{Conclusions}
	In this work, a theoretical framework is developed to analyse the dispersion of multiple charged ionic species where the flow is assumed to be a pressure-driven Poiseuille flow through the slowly varying axisymmetric microchannels under zero external electric field. The Nernst-Planck and Navie-Stokes equations are combinedly considered for the species transport where electroneutrality and zero net current is enforced to derive a self-consistent expression for the induced electric potential gradient, which arises solely from the differences in the ionic diffusivities. Through multiple-scale homogenization with small aspect ratio ($\epsilon \ll 1$), a reduced-order macrotransport equation for the cross-sectionally averaged concentrations, capturing the nonlinear coupling effect between the species via a diffusion-induced electric field. This model effectively extends the classical Taylor-Aris dispersion to the multi-species charged electrolytes, where the transport of species is governed by the interplay between the shear-enhanced advection, molecular diffusion, and geometry-modulated electro-diffusive interactions. \\
	A key findings of this work is that electroneutrality condition not only act as a passive constraint but as an active mechanism that generates an internal electric field whenever ionic mobilities differ, thereby coupling the fluxes of all species. This electro-diffusive coupling is embodied in the concentration-dependent diffusion tensor ($\mathbf{D}$), whose off-diagonal terms facilitate cross-species interactions absent in both neutral solutes and binary electrolytes. The channel geometry is found to be another key factor for species diffusion, where the mathematical model assumed the wall slope as, $\beta(x) = da/dx$ and the parameter local height $a(x)$ modulating both the advective and diffusive transport. In the case of wall converging geometry, increased velocity gradients forced the shear dispersion enhancement, but the accompanying compression strengthens electroneutrality constraints, often suppressing net spreading of different species. Conversely, diverging walls reduces the shear but can ease ion separation, allowing longer-range electro-diffusive effects to emerge.\\
	Our analysis reveals several key physical insights such as: Effective diffusivities exhibit non-monotonic, time-dependent behavior due to competition between shear-driven spreading and electro-diffusive focusing a clear departure from classical Taylor dispersion. Secondly, geometric sharpness critically influences the dispersion: triangular-wave channels, with their abrupt slope changes, generate stronger localized shear and electro-diffusive coupling than sinusoidal channels, making them more effective for rapid mixing and sharp changes in effective diffusivity. In the third observation, separation efficiency is quantified by the number of theoretical plates $N$, peaks at an intermediate arc-to-chord ratio $\Lambda \approx 1$ for convergent concave-upward channels, identifying the optimal geometry for microfluidic separations. At last, the dispersion regime depends on the Péclet number: For low $Pe$, diffusion dominates; near $Pe \approx 48$, advection and electro-diffusion balance, leading to similar effective diffusivities for all species; at high $Pe$, shear-driven dispersion prevails, though electro-diffusive coupling which modified the species diffusion and moves faster compared to earlier cases. These findings provide a predictive framework for designing microchannels that exploit wall geometry to control multi-species ionic transport without external electric fields, which can have direct applications in lab-on-a-chip separators, electrochemical sensing, and microscale mixing.\\
	The direction of future studies of the minimal model of three-species dispersion includes investigating the effects of secondary flows and Dean vortices \cite{johnson1986numerical, roberts2004shear, zhao2007effect}. We have focused only on pressure-driven flows; but, activity-driven flows \cite{saintillan2018rheology, shankar2022optimal, kumar2022transport, das2026capillary, das2024low} can also be further analyzed. While we have solved a reduced-order model, it would be fruitful to study the full system of Nernst-Planck-Poisson equations, where the breakdown of the usual electroneutrality condition can be observed at shorter time scales. \\
	\section*{Acknowledgement}
	T. M. would like to acknowledge the University Grants Commission, India (Circular No.: 221610057157), and A.K.N would like to acknowledge the Science and Engineering Research Board, India (Grant No. CRG/2023/006863) for the financial support provided during the preparation of this study.
	\section*{Declaration of Interests}
	The author reports no conflict of interest.
	\bibliographystyle{unsrt}
	\bibliography{file1}
		
\end{document}